\documentclass[aps,pre,twoside,nofootinbib,showpacs]{revtex4}
 \usepackage{amsmath,amssymb,bm,euscript}
 \usepackage{graphicx,boxedminipage}
 \usepackage[bf,small]{caption2}
 \usepackage[lflt]{floatflt}
 \usepackage{hyperref}
 \usepackage{xcolor,fancybox}
\def\WRM{Waves Random Media\ }

\definecolor{Dark-blue}{RGB}{0,0,255}
\newcommand{\bra}[1]{\ensuremath{\bm{\langle}#1\bm{|}}}
\newcommand{\ket}[1]{\ensuremath{\bm{|}#1\bm{\rangle}}}

\newcommand{\sgn}{\ensuremath\mathrm{\,sgn\,}}

\newcommand{\Av}[1]{\ensuremath{\big<#1\big>}}
\newcommand{\AV}[1]{\ensuremath{\Big<#1\Big>}}
\begin{document}

\title{THE EFFECT OF RANDOM SURFACE INHOMOGENEITIES ON MICRORESONATOR
       SPECTRAL PROPERTIES: \\THEORY AND MODELING AT MILLIMETER
       WAVE RANGE}

 \author{E.\,M. Ganapolskii}
 \author{Z.\,E. Eremenko}
 \author{Yu.\,V. Tarasov}\email[Corresponding author: ]{yutarasov@ire.kharkov.ua}
 \affiliation{Institute for Radiophysics and Electronics NASU,
 12 Proscura Street, 61085 Kharkov, Ukraine}

%------------------------------------------------------------------
\begin{abstract}
  The influence of random surface inhomogeneities on spectral
  properties of open microresonators is studied both theoretically
  and experimentally. To solve the equations governing the dynamics
  of electromagnetic fields the method of eigen-mode separation is
  applied previously developed  with reference to inhomogeneous
  systems subject to arbitrary external static potential. We prove
  theoretically that it is the gradient mechanism of wave-surface
  scattering which is the highly responsible for non-dissipative
  loss in the resonator. The influence of side-boundary
  inhomogeneities on the resonator spectrum is shown to be described
  in terms of effective renormalization of mode wave numbers jointly
  with azimuth indices in the characteristic equation. To study
  experimentally the effect of inhomogeneities on the resonator
  spectrum, the method of modeling in the millimeter wave range is
  applied. As a model object we use dielectric disc resonator (DDR)
  fitted with external inhomogeneities randomly arranged at its side
  boundary. Experimental results show good agreement with
  theoretical predictions as regards the predominance of the
  gradient scattering mechanism. It is shown theoretically and
  confirmed in the experiment that TM oscillations in the DDR are
  less affected by surface inhomogeneities than TE oscillations with
  the same azimuth indices. The DDR model chosen for our study as
  well as characteristic equations obtained thereupon enable one to
  calculate both the eigen-frequencies and the \emph{Q}-factors of
  resonance spectral lines to fairly good accuracy. The results of
  calculations agree well with obtained experimental data.
\end{abstract}
%-------------------------------------------------------------------
\pacs{05.40.-a, 02.50.Fz, 42.60.Da} %
\maketitle

%=======================================
\section{Introduction}
%=======================================

Nowadays microresonators (disk-, ring-, and spherical-shaped) evoke
considerable interest because new possibilities have recently opened
up to develop these types of resonators in the optical frequency
range and to utilize them as oscillation systems for optical lasers
\cite{bib:Polman04}. When used in lasers, such oscillation systems
offer a number of serious advantages, among which are low-threshold
currents, a high quality of the radiation spectrum,~etc.
Microresonators manufactured as a dielectric disk whose diameter is
large as compared to the wave length of the radiation are nothing
else but the open quasi-optical dielectric disk resonators which
have long been known in the resonator technology for their potential
to effectively sustain the electromagnetic (EM) field inside the
resonator volume. The retention of the field is provided due to the
total internal reflection (TIR) from the resonator side boundaries
of waves making up resonance oscillations. As a result, EM
oscillations of whispering gallery (WG) type arise. As far as their
excitation is not accompanied by the additional dissipative loss,
superhigh quality factors are achieved in the DDRs. Specifically, in
laser systems, in spite of DDR's microscopic dimensions (the disk
diameter is normally about a few $\mu$m large), the quality factors
can reach the order of $10^8$ and even more \cite{bib:Vahala03}.

The high quality factors of the DDRs, that are often prepared from
doped silicon, are generally provided not only due to extremely
small loss attainable in this material but also governed by
resonator geometry and the perfection of the crystal it is made of.
With a certain number of inhomogeneities (local and/or non-local,
random and/or regular) in the resonator material, the ray picture of
EM fields in the resonator changes dramatically as against its
perfectly homogeneous counterpart. The inhomogeneities give rise to
local violation of TIR conditions and in this way result in
additional energy loss which is evident in the quality factor drop.
From the above said there arises the problem of studying the effect
produced by random inhomogeneities in the DDRs on their spectral
characteristics.

In the theoretic analysis of the radiation loss of the DDR two types
of inhomogeneities are normally distinguished. To the first type the
inhomogeneities of volume nature belong, which are related to
regular or random spatial variations of the permittivity  in the
bulk of the material the DDR is made of. The other class of
inhomogeneities includes the so called ``surface'' imperfections
normally related to the deviation of the DDR shape from the ideal
cylindrical one. The influence of bulk random inhomogeneities on the
resonance spectrum was previously studied by the present authors in
the particular case of cavity resonators filled with randomly
distributed dielectric particles \cite{bib:GanErTar07}. It was found
that the physical mechanism through which inhomogeneities affect
resonator spectrum is basically the intermode scattering. In the
case of quasi-optical cavity resonator the peculiar feature of this
type of scattering is the selective impact of inhomogeneities on
different resonance lines. The most affected lines appear to be
those which are the least separated on the frequency axis, whereas
solitary lines are subjected to much lesser influence. Owing to such
a selective effect of random inhomogeneities, the originally dense
spectrum of the quasi-optical cavity resonator is considerably
rarefied.

In
Refs.~\cite{bib:Little97,bib:Gorod00,bib:Oraevsk2002,bib:Borselli2004},
the investigations were undertaken into the impact of \emph{surface}
inhomogeneities on spectral properties of open dielectric resonators
of cylindrical and spherical shape. The influence of boundary
roughness upon the resonance lines was described by means of the
simplified quasi-geometric approach where EM oscillations scattering
due to edge inhomogeneities is taken into account by incorporating
into the wave equation the fictitious polarization currents (PC)
randomly distributed in space \cite{bib:Kuznetsov83}. In particular,
within the framework of the model adopted in
Ref.~\cite{bib:Little97} the electromagnetic fields close to the
resonator side surface were considered as being excited by randomly
distributed near-surface current sources whose physical parameters
were phenomenologically expressed through statistical
characteristics of boundary asperities. It was precisely the
radiation produced by these sources that led to the radiation loss
of the resonator and, consequently, to the quality factor reduction.

Such an essentially phenomenological approach to the description of
the effect produced by the roughness of open resonator boundaries on
its spectral properties cannot be reckoned as satisfactory. The
volume current method suggested in Ref.~\cite{bib:Kuznetsov83},
which formed the basis for the PC concept, is, to a large extent,
rough and approximate. In this method, the radiation loss is most
frequently calculated using Green function of the Helmholtz
equation. The particular form of this function is normally chosen
proceeding from the prospective solution of wave equation in the far
wave zone. Meanwhile, the very notion of the far zone is poorly
defined for multiple sources located around the periphery of the
quasi-optical DDR we deal with in this particular work.
Specifically, when deriving characteristic equations for open DDR
eigen-frequencies, one needs to join EM field components exactly at
the boundaries of the system under consideration. At this point a
significant uncertainty can arise because the actual external fields
subject to \emph{local} joining with the internal ones in the
presence of surface roughness can deviate considerably from the
fields approximated into the boundary vicinity from large distances,
i.\,e. from the far wave zone.

To correctly determine the EM fields near the random-inhomogeneous
resonator surface the theories specially adapted for the description
of classic and/or quantum wave scattering by rough interfaces should
be applied (see, e.\,g.,
Refs.~\cite{bib:BassFuks79,bib:Ogilvy91,bib:Voronovich94}). These
theories are generally applicable to the cases where wave scattering
by random rough surfaces is in a sense weak. Typically, this implies
fluctuations of system boundaries to be relatively small in height
and sufficiently smooth, so that the applicability of Rayleigh
hypothesis \cite{bib:Rayleigh1907,bib:Rayleigh45} was not violated.
However, for confined systems like the DDR considered in this work
the issues pertinent to wave scattering by surface inhomogeneities
appear to be much more complicated. Firstly, in practice the
inhomogeneities are not always small enough, as well as sufficiently
smooth, therefore the conditions for scattering weakness are easily
violated. Moreover, as the resonance ray trajectories in
high-\emph{Q} resonators are periodic, the effect of oscillation
scattering caused by the boundary roughness is ``path''-accumulated.
Scattering can become strong even though the asperities are
small-in-height and smooth. This makes the applicability of the
above-mentioned theories of rough-surface scattering in the case of
high-quality resonators highly questionable.

In Refs.~\cite{bib:MakTar98,bib:MakTar01}, the novel transport
theory was developed for waveguide-shaped systems with random rough
boundaries. In the framework of this theory it was revealed that the
wave scattering resulting from fluctuations of inter-media surfaces
can be efficiently described in terms of two physical mechanisms,
namely, the \emph{amplitude} and the \emph{gradient} ones. For the
first mechanism it is just the mean-square height of the asperities
that serves as a main guiding parameter, whereas for the second one
the mean slope of the asperities or, in other words, their
sharpness, plays the decisive role. Both of these mechanisms
contribute additively to the scattering amplitude, but partial
probabilities pertaining to them may differ essentially. It was
shown in \cite{bib:MakTar98,bib:MakTar01} that, in most cases, the
role of the gradient scattering appears to be prevalent. Yet there
have been no experimental confirmations of this fact in the
literature so far.

Apart from peculiar problems associated with surface nature of
scattering in edge-disordered resonators, some more questions arise
which need to be resolved when studying the spectra of DDRs subject
to surface inhomogeneities. Among these questions, particularly, are
those related to the vector nature of fields which are excited in
these essentially non-integrable systems. It is known that even for
ideal cylindrical DDRs the electrical- and magnetic-type
oscillations cannot be separated in the strict sense. They always
remain to the certain extent intermixed \cite{bib:Vainstein88}. This
fact renders theoretic analysis of such systems spectra rather
sophisticated. Spectral properties of DDRs without random
inhomogeneities were examined in a number of theoretical papers
(see, e.\,g.,
Refs.~\cite{bib:IvanovKalin1988,bib:Peng96,bib:Anino97}). Yet, in
deriving characteristic equations some inexact \emph{a priori}
assumptions where used, which still require both theoretical
grounding and experimental corroboration.

In our present study, one of the goals is to investigate
theoretically the physical mechanisms responsible for widening
resonance lines of dielectric microresonators with random surface
inhomogeneities. In view of such systems being non-integrable, it is
necessary to elaborate an appropriate theoretical model allowing for
sufficiently accurate determination of both the frequency spectrum
and the quality factors of resonance lines. Yet another goal of this
study is to corroborate experimentally which of the physical
mechanisms does play a dominant role in the scattering of EM
oscillations excited in microresonators with random rough side
boundaries.

From the theoretical viewpoint, the spectrum analysis in our study
is carried out in terms of scalar potentials, specifically,
electrical and magnetic Hertz functions~\cite{bib:Vainstein88}. We
formulate the conditions wherein these potentials make independent
contributions to EM fields in the resonator, which is equivalent to
decoupling the oscillations of TE and TM polarization. The Helmholtz
equation for Hertz potentials in an irregular-shaped open resonator
is equivalent to Schr\"odinger equation for electrons moving in the
piecewise continuous space subject to random potential. Owing to
this one can extend the results obtained in the present study to
quantum systems as well, in particular, to open quantum dots having
random rough boundaries.

To solve wave equations in the rough-bounded resonator, the
eigen-mode separation method is used, previously developed with
reference to waveguide-like systems subject to arbitrary static
potential \cite{bib:Tar00,bib:Tar03,bib:Tar05}). Using this method,
the originally posed statistical problem of determining the fields
in three-dimensional DDR with complex, randomly rough, side boundary
is rigorously reduced to the set of one-dimensional dynamic
equations that contain some \emph{effective} random potentials. We
show that under the conditions where gradient scattering mechanism
is dominant, the boundary roughness effect on the resonator spectrum
can be described through certain renormalization of mode wave
numbers and azimuth indices of the Bessel functions in the
characteristic equation. The values of renormalized wave numbers and
mode indices decrease as the asperities get sharpened, which is
consistent with a decrease in the resonator \emph{Q}-factor. We are
thus led to the conclusion that the observed reduction of resonant
line quality factors results not from the extra dissipative loss but
rather from EM field intermode Rayleigh scattering induced by random
surface inhomogeneities. The imperfection of the resonator shape
results in the local violation of the TIR conditions. This leads to
additional radiation loss of EM energy and, hence, to a decrease in
the level of localization of EM fields inside the resonator.

Since characteristic dimensions of surface inhomogeneities in actual
microresonators are always quite small (of the order of nanometers),
to verify our theoretical findings experimentally we have decided
upon the method of simulation with macroscopic devices. As a model
system we have employed a millimeter wave quasi-optical resonator
made of a circular teflon disk. WG oscillations of TE and TM types
were excited in the disk using a special waveguide antenna. The
inhomogeneities of the resonator side boundary were made in the form
of teflon bracket-bars randomly attached to the outside cylindrical
surface. Our experimental results have demonstrated excellent
qualitative agreement with the developed theory as regards spectral
lines widening caused by the resonator side boundary roughness.
Furthermore, the relatively simple model of the DDR field
distribution, which was adopted in our study, enabled us to
calculate both the \emph{Q}-factors and the frequencies of the
resonance lines with quite satisfactory accuracy. The calculations
appeared to be in fair conformity with our experimental data.

%=======================================
\section{Theoretical model and derivation of basic equations}
%=======================================

\begin{figure}[h!!!]
  \setcaptionmargin{.8in}%
  \centering
  \scalebox{.9}[.9]{\includegraphics{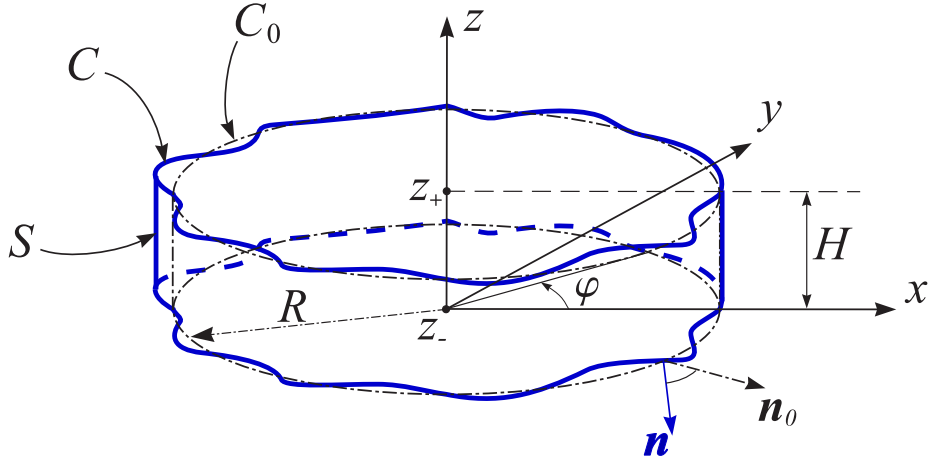}}
%\captionstyle{hang}
  \caption{(Color online) The sketch of the dielectric cylindrical resonator
  with a randomly rough side wall. Vectors $\mathbf{n}$ and
  $\mathbf{n_0}$ point out the local normal directions  to the rough ($S$)
  and to the smoothed resonator side surface, respectively.
  \hfill\label{fig1}}
\end{figure}

Consider an open disk resonator as a finite-height cylinder made of
the dielectric material with permittivity~$\varepsilon_0$.
Plain-parallel end boundaries of the cylinder traverse the central
axis ($z$) at points $z_{\pm}=\pm H/2$ (see Fig.~\ref{fig1}), the
side boundary ($S$) is formed by the generatrix passing parallel to
$z$-axis along closed contour $C$
whose distance from the central axis is given by
\begin{equation}\label{SideBound}
  r(\varphi)=R+\xi(\varphi)\ ,\qquad\qquad \varphi\in[-\pi,\pi]\ ,
\end{equation}
$R$ is the radius of the averaged (i.\,e., ideally circular) contour
$C_0$ containing no random bends. The function $\xi(\varphi)$ will
be regarded as the Gaussian random process with a zero mean value,
$\big<\xi(\varphi)\big>=0$, and binary correlation function
\begin{equation}\label{BinCorr}
  \big<\xi(\varphi)\xi(\varphi')\big>=\sigma^2W(\varphi-\varphi')\ .
\end{equation}
Here, $\sigma$ is the mean-square height of boundary asperities,
$W(\varphi)$ is the dimensionless function which has a unit maximal
value at zero argument and falls to parametrically small values at
angle distance $\varphi_c\ll 1$ (the correlation angle). The angle
brackets in Eq.~\eqref{BinCorr} denote statistical averaging over
the ensemble of realizations of random function $\xi(\varphi)$. In
what follows, for the estimation purposes, along with angle
parameter $\varphi_c$ we will use another correlation parameter,
viz., arc correlation length $s_c=\varphi_cR$. Both the random
function $\xi(\varphi)$ and the regular $W(\varphi)$ will be thought
of as periodic with period~$2\pi$.

It is well known that in general the EM field of an open dielectric
resonator can be given as a superposition of oscillations of
electric and magnetic types (TM and TE polarized, respectively)
\cite{bib:Vainstein88}. The vector fields of both of these types can
be expressed in terms of scalar potentials, $U(\mathbf{r})$ and
$V(\mathbf{r})$ (the so-called electric and magnetic Debye
potentials), which meet the same Helmholtz equation but are subject
to different joining conditions at the resonator boundaries. In
arbitrarily shaped resonators TM and TE oscillations are essentially
intermixed, being strictly decoupled only in the case of
sufficiently symmetric systems such as, e.\,g., an infinite
dielectric cylinder \cite{bib:Vainstein88} or a dielectric sphere
\cite{bib:Oraevsk2002}. Below it will be shown that TM and TE
oscillations can also be decoupled (with high accuracy, though
approximately) in the case of the cylindrical DDR of arbitrary
thickness. This fact opens an opportunity to extend the conclusions
of the present work to electronic microresonance systems, e.\,g.,
partially open quantum dots.

In studying oscillations in the DDR with randomly rough side
boundaries we first reduce the problem of wave scattering at the
boundary to the problem of scattering in the bulk of the resonator
of ideal circular form. The Helmholtz equation for scalar wave field
$\Psi(\mathbf{r})$, whose role is played by one of the
above-mentioned potentials, after rewriting it in cylindrical
coordinates and using the conformal coordinate transformation,
\begin{eqnarray}
    \widetilde{r} &=& \frac{r}{1+\xi(\varphi)/R}\ ,\notag \\
\label{coord_transform}
    \widetilde{\varphi} &=& \varphi \ ,\\\notag
    \widetilde{z} &=& z\ ,
\end{eqnarray}
is reduced to the form
\begin{equation}\label{GreenEqMain}
  \left[\frac{1}{r}\frac{\partial}{\partial r}r
  \frac{\partial}{\partial r}+
  \frac{1}{r^2}\frac{\partial^2}{\partial\varphi^2}+
  \frac{\partial^2}{\partial z^2}+K^2(r,z)
  -\hat{V}^{(h)}-\hat{V}^{(s)}\right]
  \Psi(r,\varphi,z)=0
\end{equation}
(the tilde signs over coordinate variables from here on are
omitted). Here, $K^2(r,z)=k^2\varepsilon(r,z)$,
\begin{equation}\label{Epsilon(r,z)}
  \varepsilon(r,z)=\left\{
\begin{array}{cl}
  \varepsilon_0+i/\tau_0&,\qquad (r,z)\in\Omega \\[6pt]
  1&,\qquad (r,z)\not\in\Omega
\end{array}\right.\ ,
\end{equation}
$1/\tau_0$ is the phenomenological frequency parameter which takes
into account dissipative and other uncontrollable losses in the
system, except for the radiation loss, $\Omega$ is the bulk region
occupied by the dielectric. Effective potentials $\hat{V}^{(h)}$ and
$\hat{V}^{(s)}$ in Eq.~\eqref{GreenEqMain} are the operators whose
coordinate representation reads
\begin{subequations}\label{Potentials}
\begin{align}\label{heightPot}
  & \hat{V}^{(h)} = -\left[K^2(r,z)+\frac{\partial^2}{\partial
  z^2}\right]
  \left[\beta^2(\varphi)-1\right]\ ,\\
\label{slopePot}
  & \hat{V}^{(s)} = \left[\frac{\xi'(\varphi)}{R\beta(\varphi)}
  \frac{\partial}{\partial\varphi}+\frac{\partial}{\partial\varphi}
  \frac{\xi'(\varphi)}{R\beta(\varphi)}\right]\frac{1}{r}\frac{\partial}{\partial r}
  -\left[\frac{\xi'(\varphi)}{R\beta(\varphi)}\right]^2
  \frac{1}{r}\frac{\partial}{\partial r}r\frac{\partial}{\partial r}\
  ,\\[6pt]
\notag
  & \beta(\varphi)=1+\xi(\varphi)/R\ .
\end{align}
\end{subequations}
Indices ``$h$'' and ``$s$'' specifying potentials \eqref{Potentials}
indicate that the corresponding potential is mainly governed either
by fluctuations in the asperity height (i.\,e. by height function
$\xi(\varphi)$) or by fluctuations in the asperity slope (i.\,e., by
slope function $\xi'(\varphi)$). Such a subdivision of the
potentials, with regard to the asymptotic suppression of
correlations between functions $\xi(\varphi)$ and $\xi'(\varphi)$
over angle intervals that are large as compared to the correlation
angle, urges one to consider potentials \eqref{heightPot} and
\eqref{slopePot} as corresponding to different physical wave-surface
scattering mechanisms~\cite{bib:MakTar98,bib:MakTar01}. We will
below refer to these mechanisms as the amplitude (or the height,
\emph{h}) and the gradient (or the slope, \emph{s}) scattering
mechanisms, respectively.

Subsequently we will examine the boundary asperities which are
sufficiently small in hight so as to meet inequality
\begin{equation}\label{SmallHeight}
  \sigma\ll R\ .
\end{equation}
In contrast to the widespread belief (see, e.\,g.,
Ref.~\cite{bib:BassFuks79}), this does not necessarily imply that
the surface-roughness-induced scattering be regarded as weak. We
note, for example, that the local value of the potential
\eqref{slopePot} is estimated by the parameter $\sigma/s_c$, which
may take an arbitrary absolute value provided the condition
\eqref{SmallHeight} is met. The true conditions for the scattering
to be classified as weak will be provided below, based on the
operator technique applied to perform the principal calculations.

%=======================================
\section{Separation of azimuth modes in randomly rough disk resonator}
\label{Mode_Separation}
%=======================================

The potentials $\hat{V}^{(h)}$ and $\hat{V}^{(s)}$ in
Eq.~\eqref{GreenEqMain}, that account for inhomogeneity of the
resonator side-boundary, are defined not quite conveniently from the
viewpoint of subsequent use of perturbation theories. The
inconvenience relates to non-zero average value of the potential
\eqref{slopePot}. By separating this average, we can rewrite
Eq.~\eqref{GreenEqMain} as
\begin{equation}\label{PsiEq-alt}
  \bigg[(1+\Xi^2)\frac{1}{r}\frac{\partial}{\partial r}r
  \frac{\partial}{\partial r}+
  \frac{1}{r^2}\frac{\partial^2}{\partial\varphi^2}+
  \frac{\partial^2}{\partial z^2}+K^2(r,z)
  -\hat{V}^{(h)}-\hat{V}^{(s1)}-\hat{V}^{(s2)}\bigg]
  \psi(r,\varphi,z)=0\ ,
\end{equation}
where two different slope potentials are introduced instead of
potential Eq.~\eqref{slopePot}, which have zero mean values, viz.

\begin{subequations}\label{SlopePots-defs}
\begin{align}
 \label{SlPot-1}
  \hat{V}^{(s1)} =& \frac{1}{R}\left[\xi'(\varphi)
  \frac{\partial}{\partial\varphi}+\frac{\partial}{\partial\varphi}
  \xi'(\varphi)\right]\frac{1}{r}\frac{\partial}{\partial r}\ ,\\
 \label{SlPot-2}
  \hat{V}^{(s2)} =& -\left[{\xi'}^2(\varphi)/R^2-\Xi^2\right]
  \frac{1}{r}\frac{\partial}{\partial r}r\frac{\partial}{\partial
  r}\ .
\end{align}
\end{subequations}
The parameter $\Xi$ in Eqs.~\eqref{PsiEq-alt} and \eqref{SlPot-2},
which is defined as
\begin{equation}\label{Xi-def}
  \Xi^2=\frac{1}{R^2}
  \Av{\left[\xi'(\varphi)\right]^2}
  \sim\frac{\sigma^2}{s_c^2}\ ,
\end{equation}
serves as a measure for the mean-square slope of boundary
asperities.

Considering that irregularities of the resonator side surface are
aligned paraxially, they cannot result in additional scattering
along axis $z$ except for the partial reflection produced by the
initial step change in this direction of the permittivity of the
medium. This makes it possible to remove from Eq.~\eqref{PsiEq-alt}
the dependence on the coordinate $z$ with the model method similar
to the one we use when analyzing the spectrum of homogeneity-free
DDR (see Appendix~\ref{Unperturbed_spectrum}). Specifically, we
assume the dependence of the mode wave function on the coordinate
$z$ to have the same form used to find $s$- and $a$-solutions for
the ideal-shaped resonator (for the particular $E_z$-symmetric
solution this dependence is given in Eqs.~\eqref{Un(s)} and
\eqref{Vn(s)}). Afterwards, equation~\eqref{PsiEq-alt}, being
divided by the factor of $1+\Xi^2$, is reduced to the following
form,
\begin{equation}\label{PsiEq-alt-renorm}
  \bigg[\frac{1}{r}\frac{\partial}{\partial r}r
  \frac{\partial}{\partial r}+
  \frac{1}{(1+\Xi^2)r^2}\frac{\partial^2}{\partial\varphi^2}+
  \widetilde{K}_{\bot}^2(r|z_{\scriptscriptstyle\lessgtr})
  -\widetilde{{V}}^{(h)}(z_{\scriptscriptstyle\lessgtr})-\widetilde{\hat{V}}^{(s1)}-
  \widetilde{\hat{V}}^{(s2)}\bigg]
  \Psi(r,\varphi|z_{\scriptscriptstyle\lessgtr})=0\ ,
\end{equation}
where
\begin{subequations}\label{Renormed_parameters}
\begin{alignat}{2}
 \label{K-renorm}
   \widetilde{K}_{\perp}^2(r|z_{\scriptscriptstyle\lessgtr})&=
   \frac{1}{1+\Xi^2}\left\{
    \begin{array}{clcl}
     \big[\varepsilon\theta(R-r)+\theta(r-R)\big]k^2-k_z^2  &\ \text{,}&\qquad |z|&<H/2 \\
     k^2+\varkappa_z^2  &\ \text{,}&\qquad |z|&>H/2
    \end{array}\ ,\right.\\
 \label{V(h)-renorm}
   \widetilde{{V}}^{(h)}(z_{\scriptscriptstyle\lessgtr})&=
   -\widetilde{K}_{\perp}^2(r|z_{\scriptscriptstyle\lessgtr})\left[\beta^2(\varphi)-1\right]\
   .
\end{alignat}
\end{subequations}
The tilde signs over the potentials in Eq.~\eqref{PsiEq-alt-renorm}
denote their renormalization by the factor of
$\big(1+\Xi^2\big)^{-1}$. Equation \eqref{PsiEq-alt-renorm} actually
describes two-dimensional fields because symbol
$z_{\scriptscriptstyle\lessgtr}$ acts not as the current axial
coordinate but stands for some axial index specifying in which
domain of the $z$-axis --- either $|z|<H/2$ or $|z|>H/2$
--- the solution to Eq.~\eqref{PsiEq-alt} is sought.

In the azimuth mode representation, equation
\eqref{PsiEq-alt-renorm} assumes the form
\begin{equation}\label{Psi-Azim_repr}
  \left[\frac{1}{r}\frac{\partial}{\partial r}r\frac{\partial}{\partial r}
  +\widetilde{K}_{\perp}^2(r|z_{\scriptscriptstyle\lessgtr})-\frac{\widetilde{n}^2}{r^2}
  -\widetilde{\mathcal{V}}_{n}(r|z_{\scriptscriptstyle\lessgtr})\right]
  \Psi_{n}(r|z_{\scriptscriptstyle\lessgtr})-\sum_{m\neq n}
  \widetilde{\mathcal{U}}_{nm}(r|z_{\scriptscriptstyle\lessgtr})\Psi_{m}(r|z_{\scriptscriptstyle\lessgtr})=0\ .
\end{equation}
Here,
\begin{equation}\label{U_nm-def}
  \widetilde{\mathcal{U}}_{nm}(r|z_{\scriptscriptstyle\lessgtr})=\oint
  d\varphi\bra{\varphi,n}\widetilde{V}(r,\varphi|z_{\scriptscriptstyle\lessgtr})\ket{\varphi,m}\
\end{equation}
is the matrix element of the entire potential
$\widetilde{V}(r,\varphi|z_{\scriptscriptstyle\lessgtr})=
\widetilde{{V}}^{(h)}(z_{\scriptscriptstyle\lessgtr})+\widetilde{\hat{V}}^{(s1)}+
\widetilde{\hat{V}}^{(s2)}$, which is taken between eigen-functions
of the azimuth part of the Laplace operator (see
Appendix~\ref{Unperturbed_spectrum}),
$\widetilde{\mathcal{V}}_{n}(r|z_{\scriptscriptstyle\lessgtr})
\equiv\widetilde{\mathcal{U}}_{nn}(r|z_{\scriptscriptstyle\lessgtr})$,
the index $n$ changes to ${\widetilde{n}=n\big/\sqrt{1+\Xi^2}}$.
Later we will refer to matrix elements $\widetilde{\mathcal{V}}_{n}$
and $\widetilde{\mathcal{U}}_{nm}$ as the intra- and inter-mode
potentials, respectively.

In the general case it is rather difficult to immediately solve the
infinite set of coupled equations \eqref{Psi-Azim_repr}. However,
the solution can be obtained in terms of the operator technique
previously developed by the present authors with reference to
waveguide-type random systems of arbitrary dimensionality
\cite{bib:Tar00,bib:Tar03} and advantageously applied afterwards to
the analysis of bulk-disordered cavity resonators
\cite{bib:GanErTar07}. The above technique, as applied to the
problem touched upon in the present paper, is adapted in
Appendix~\ref{Mode_separation}. The advantage of the technique is
that it can be used to derive precise closed equations for wave
functions of each of the azimuth modes, namely,
\begin{equation}\label{Psi_n-separate}
  \left[\frac{1}{r}\frac{\partial}{\partial
  r}r\frac{\partial}{\partial r}
  +\widetilde{K}_{\perp}^2(r|z_{\scriptscriptstyle\lessgtr})-\frac{\widetilde{n}^2}{r^2}
  -\widetilde{\mathcal{V}}_{n}(r|z_{\scriptscriptstyle\lessgtr})-
  \hat{\mathcal T}_n\right]
  \Psi_{n}(r|z_{\scriptscriptstyle\lessgtr})=0\ .
\end{equation}
Here, along with local intra-mode potential
$\widetilde{\mathcal{V}}_{n}$, the operator potential $\hat{\mathcal
T}_n$ arises, which rigorously allows for the inter-mode scattering.
The structure of this potential, though well-recognized, is quite
complicated to be operated with at an arbitrary scattering
intensity. Yet, the estimations we provide in the next section
substantially simplify the \emph{T}-potential in
Eq.~\eqref{Psi_n-separate} in different limiting cases, and in this
way they allow one to obtain the oscillation spectrum of a randomly
rough DDR in almost all physically sensible situations.

%=======================================
\section{Spectrum of the DDR with weakly rough side boundary}
\label{Rough_Specrum}
%=======================================

Hereinafter we will refer to the system as weakly rough one if
r.m.s. height of its boundary asperities meets
inequality~\eqref{SmallHeight}. As the potentials in
Eq.~\eqref{Psi_n-separate} are of the operator nature, we will
estimate their strength using standard definition of the operator
norm~\cite{bib:KolmFom68,bib:Kato66}. For our purposes the formula
\begin{equation}\label{norm_def}
  \Av{\big\|\hat{\mathsf A}\big\|^2}=\sup_{0\neq\psi\in\mathbb{X}}
  \frac{\Av{(\hat{\mathsf A}\psi,\hat{\mathsf A}\psi)}}{(\psi,\psi)}
\end{equation}
is the most appropriate, in which the parenthesis symbolize the
scalar product on the functional space $\mathbb{X}$ consisting of
the class of solutions to Eq.~\eqref{Psi_n-separate} with no random
potentials.

To estimate the potentials in Eq.~\eqref{PsiEq-alt-renorm} we first
calculate their azimuth matrix elements. For relatively small-height
asperities the ``height'' potential ${\widetilde{V}}^{(h)}$ equals
approximately
${\widetilde{V}}^{(h)}(z_{\scriptscriptstyle\lessgtr})\approx
-2\widetilde{K}_{\perp}^2(r|z_{\scriptscriptstyle\lessgtr})\xi(\varphi)/R$.
Mode matrix elements of this potential are
\begin{equation}\label{Unm(h)}
  \widetilde{\mathcal{U}}_{nm}^{(h)}(r|z_{\scriptscriptstyle\lessgtr})=
  -\sqrt{\frac{2}{\pi}}\widetilde{K}_{\perp}^2(r|z_{\scriptscriptstyle
  \lessgtr})\frac{\widetilde{\xi}(n-m)}{R}\
  ,
\end{equation}
$\widetilde{\xi}(n)$ is the Fourier transform of the function
$\xi(\varphi)$. Matrix elements of the ``slope'' potentials
$\widetilde{\hat{V}}^{(s1)}$ and $\widetilde{\hat{V}}^{(s2)}$ are
equal, respectively,
\begin{subequations}\label{Unm-alt}
\begin{align}
 \label{=Unm(s1)-alt}
 & \widetilde{\mathcal{U}}_{nm}^{(s1)}(r)=
  -\frac{1}{\sqrt{2\pi}(1+\Xi^2)R}\left(n^2-m^2\right)\widetilde{\xi}(n-m)
  \frac{1}{r}\frac{\partial}{\partial r}\ ,\\
 \label{=Unm(s2)-alt}
 & \widetilde{\mathcal{U}}_{nm}^{(s2)}(r)=
  \frac{1}{2\pi(1+\Xi^2) R^2}\sum_{l=-\infty}^{\infty}
  (n-l)(l-m)\left[\widetilde{\xi}(n-l)\widetilde{\xi}(l-m)-
  \AV{\widetilde{\xi}(n-l)\widetilde{\xi}(l-m)}\right]
  \frac{1}{r}\frac{\partial}{\partial r}r\frac{\partial}{\partial
  r}\ .
\end{align}
\end{subequations}
%

%=======================================
\subsection{The efficiency of intra-mode scattering}
%=======================================

We will ignore the role of the potential \eqref{Unm(h)} for the
intra-mode scattering proceeding from the following considerations.
It~is evident that under condition \eqref{SmallHeight} the uniform
azimuth mode of this potential can result, at most, in relatively
small ($\sim \sigma/R$) renormalization of the unperturbed
``in-plane energy'' $\widetilde{K}_{\perp}^2(r|z)$. Yet, in the case
we consider below even this does not occur. We will regard the
asperities as not only being small in height but also as
small-scaled ones in the sense that the inequality holds
\begin{equation}\label{Small-scale}
  s_c\ll R\quad\Longleftrightarrow\quad\varphi_c\ll 1\ .
\end{equation}
In this limit the uniform mode of the random process $\xi(\varphi)$
with parametric accuracy
\big($\widetilde{\xi}(0)\sim\sigma\varphi_c$\big) can be put equal
to zero, since the process is nearly ergodic within the interval of
the angle variable change.

Gradient potential \eqref{=Unm(s1)-alt} is not involved in the
intra-mode scattering by its definition, so that only the potential
\eqref{=Unm(s2)-alt} should be taken into account, where one must
let $m=n$. We use as trial functions in Eq.~\eqref{norm_def} the
solutions of Eq.~\eqref{Psi_n-separate} with random potentials equal
to zero, i.\,e., Bessel function
$J_{\widetilde{n}}\big[\widetilde{K}_{\perp}(r|z_{\scriptscriptstyle\lessgtr})r\big]$
in the interval $0<r<R$ and Hankel function
$H^{(1)}_{\widetilde{n}}\big[\widetilde{K}_{\perp}(r|z_{\scriptscriptstyle\lessgtr})r\big]$
in the domain $R<r<\infty$. Such a choice is justified provided the
intra-mode scattering has a~slight impact on the mode energies. Then
we arrive at the following estimate for the
potential~$\widetilde{{\mathcal{V}}}_{n}$ norm,
\begin{equation}\label{Vn-norm_estim}
  \Av{\big\|\widetilde{{\mathcal{V}}}_{n}\big\|^2}\sim
  \frac{\widetilde{K}_{\perp}^4}{(1+\Xi^2)^2R^4}\AV{\bigg[
  \sum_{l=-\infty}^{\infty}l^2\Big(\big|\widetilde{\xi}(l)\big|^2-
  \Av{\big|\widetilde{\xi}(l)\big|^2}\Big)\bigg]^2}\ .
\end{equation}
After averaging Eq.~\eqref{Vn-norm_estim} using correlation equality
\begin{equation}\label{BinCorr-Fourier}
  \Av{\widetilde{\xi}(k)\widetilde{\xi}^*(l)}=
  \sqrt{2\pi}\sigma^2\widetilde{W}(k)\delta_{kl}\ ,
\end{equation}
which immediately stems from Eq.~\eqref{BinCorr} ($\widetilde{W}$ is
the Fourier transform of correlation function $W(\varphi)$), we
obtain
\begin{equation}\label{Vn-norm_estim-fin}
  \Av{\big\|\widetilde{{\mathcal{V}}}_{n}\big\|^2}\sim
  \frac{\widetilde{K}_{\perp}^4}{(1+\Xi^2)^2}\left(
  \frac{\sigma}{R}\right)^4\sum_{l=-\infty}^{\infty}
  l^4\widetilde{W}^2(l)\approx\widetilde{K}_{\perp}^4\frac{\sigma}{R}
  \frac{\big(\sigma/s_c\big)^3}{\big(1+\sigma^2/s_c^2\big)^2}\ .
\end{equation}
It can be readily seen that the factor standing at
$\widetilde{K}_{\perp}^4$ in the right-hand side of
Eq.~\eqref{Vn-norm_estim-fin} is, in view of
Eq.~\eqref{SmallHeight}, small as compared to unity. This
substantiates the above assumption about the intra-mode scattering
weakness.

%=======================================
\subsection{Comparative estimations of inter-mode potentials}
%=======================================

The inter-mode scattering rate in our resonator is up to the norm of
the operator $\hat{\mathsf{R}}$ entering \emph{T}-matrix
\eqref{T_oper-suppl}.

\emph{1. The ``amplitude'' inter-mode scattering.}

When estimating the norm of the ``height'' item
$\hat{\mathsf{R}}^{(h)}$ in the operator $\hat{\mathsf{R}}$ one is
faced with a need to evaluate the expression
\begin{equation}\label{R(h)_estim-alt}
  \Av{\big\|\hat{\mathsf{R}}^{(h)}\big\|^2}\sim
  \frac{1}{\|\psi\|^2}\sum_k\int\limits_0^{\infty}rdr
  \sum_{m_1,m_2\neq
  k}\int\limits_0^{\infty}r_1dr_1\int\limits_0^{\infty}r_2dr_2
  G^{(V)}_{k}(r,r_1){G^{(V)}_{k}}^*(r,r_2)
  \AV{\widetilde{\mathcal{U}}_{km_1}^{(h)}(r_1){\widetilde{\mathcal{U}}_{km_2}^{(h)*}}(r_2)}
  \psi_{m_1}(r_1)\psi_{m_2}^*(r_2)
\end{equation}
(in order not to overload subsequent formulae we omit axial index
$z_{\scriptscriptstyle\lessgtr}$, as this cannot lead to
misunderstanding; function $G^{(V)}_{k}(r,r_1)$ is introduced in
Appendix~\ref{Mode_separation}). The simultaneous presence in
Eq.~\eqref{R(h)_estim-alt} of both the integrals over radial
coordinate $r$ and the sums over mode indices stems from the
definition of the functional space where the operator
$\hat{\mathsf{R}}$ is effective.

In view of Eqs.~\eqref{Unm(h)} and \eqref{BinCorr-Fourier} the
correlator in the integrand of Eq.~\eqref{R(h)_estim-alt} is
calculated to
\begin{equation}\label{CorrUkm1Ukm2}
\AV{\widetilde{\mathcal{U}}_{km_1}^{(h)}(r_1){\widetilde{\mathcal{U}}_{km_2}^{(h)*}}(r_2)}=
  2\sqrt{\frac{2}{\pi}}\left(\frac{\sigma}{R}\right)^2
  \widetilde{K}^2_{\perp}(r_1)\widetilde{K}^{2*}_{\perp}(r_2)
  \widetilde{W}(k-m_1)\delta_{m_1m_2}\ .
\end{equation}
By substituting this into Eq.~\eqref{R(h)_estim-alt} and assuming
the asperity correlation function to have the Gaussian form, viz.
${W(\varphi)=\exp\big(-\varphi^2/2\varphi_c^2\big)}$, we arrive at
the following estimate for the height term in the inter-mode
scattering operator,
\begin{equation}\label{R(h)_estim-alt(fin)}
  \Av{\big\|\hat{\mathsf{R}}^{(h)}\big\|^2}\sim
  (\widetilde{K}_{\perp}\sigma)^2\ .
\end{equation}
The Rayleigh parameter $k\sigma$, to whose square the right-hand
side of Eq.~\eqref{R(h)_estim-alt(fin)} is proportional, may take
small or large values, depending upon the relationship between
mean-square height of the asperities and the wave length of the
excited oscillations. According to the value of this parameter we
will distinguish between weak ($k\sigma\ll 1$) and strong
($k\sigma\gg 1$) inter-mode scattering caused by the ``height''
potential.

\emph{2. The ``gradient'' inter-mode scattering.}

This type of scattering between different azimuth modes is related
to the availability in Eq.~\eqref{Psi-Azim_repr} of the
potentials~\eqref{Unm-alt}. Their correlators needed to estimate the
operator norms of items $\hat{\mathsf{R}}^{(s1)}$ and
$\hat{\mathsf{R}}^{(s2)}$ in the operator $\hat{\mathsf{R}}$, are
equal, respectively,
\begin{subequations}\label{CorrUkm1Ukm2_(ab)}
\begin{align}\label{CorrUkm1Ukm2(s1)}
  \AV{\widetilde{\mathcal{U}}_{km_1}^{(s1)}(r_1)
  {\widetilde{\mathcal{U}}_{km_2}^{(s1)*}}(r_2)} =&
  \frac{1}{\sqrt{2\pi}}\frac{\sigma^2}{(1+\Xi^2)^2R^2}(k^2-m_1^2)^2\widetilde{W}(k-m_1)\delta_{m_1m_2}
  \frac{1}{r_1}\frac{\partial}{\partial r_1}\cdot\frac{1}{r_2}\frac{\partial}{\partial r_2} \
  ,\\[12pt]
    \AV{\widetilde{\mathcal{U}}_{km_1}^{(s2)}(r_1){\widetilde{\mathcal{U}}_{km_2}^{(s2)*}}(r_2)}=&
  \frac{1}{(2\pi)^2(1+\Xi^2)^2R^4}
   \left(\frac{1}{r_1}\frac{\partial}{\partial r_1}r_1\frac{\partial}{\partial r_1}\right)
  \left(\frac{1}{r_2}\frac{\partial}{\partial r_2}r_2\frac{\partial}{\partial r_2}\right)
    \notag\\
\label{CorrUkm1Ukm2(s2)}
  &\times
  \sum_{l_1,l_2=-\infty}^{\infty}
  (k-l_1)(l_1-m_1)(k-l_2)(l_2-m_2)
  \AV{\widetilde{\xi}(k-l_1)\widetilde{\xi}(l_1-m_1)
  \widetilde{\xi}^*(k-l_2)\widetilde{\xi}^*(l_2-m_2)}
   \ .
\end{align}
\end{subequations}
Assuming the function $\xi(\varphi)$ to be the Gaussian random
process, the equality \eqref{CorrUkm1Ukm2(s2)} can be partially
simplified as the double sum over $l_1$ and $l_2$ is reduced to the
single one,
\begin{equation}\label{Sum_l_1l_2-result}
  \sum_{l_1,l_2=-\infty}^{\infty}\ldots\ =
  4\pi\sigma^4\delta_{m_1m_2}\sum_{l=-\infty}^{\infty}
  (k-l)^2(l-m_1)^2\widetilde{W}(k-l)\widetilde{W}(l-m_1)\ .
\end{equation}
Then the alternate substitution of correlation functions
\eqref{CorrUkm1Ukm2_(ab)} into Eq.~\eqref{R(h)_estim-alt} in place
of the height potential correlator results in the following
estimation formulas for operators $\hat{\mathsf{R}}^{(s1)}$ and
$\hat{\mathsf{R}}^{(s2)}$,
\begin{subequations}\label{R(s1s2)_nrm-estim}
 \begin{align}
 \label{R(s1)_nrm-estim}
   \Av{\big\|\hat{\mathsf{R}}^{(s1)}\big\|^2} &\sim
   \frac{\Xi^2}{(1+\Xi^2)^2}\cdot\frac{1}{\varphi_c^2} \ ,\\[6pt]
   \label{R(s2)_nrm-estim}
   \Av{\big\|\hat{\mathsf{R}}^{(s2)}\big\|^2} &\sim
   \left(\frac{\Xi^2}{1+\Xi^2}\right)^2(\widetilde{K}_{\perp}R)^2\ .
 \end{align}
\end{subequations}

The collation of norms \eqref{R(h)_estim-alt(fin)} and
\eqref{R(s1s2)_nrm-estim} reveals that the height and the gradient
scattering mechanisms, given their effect is evaluated as a function
of roughness statistical parameters and radial components of the
azimuth mode wave vectors, can essentially compete against one
another. However, the general statement boils down to the fact that
it is the gradient scattering which is prevalent in the most part of
the parameter region. The amplitude scattering mechanism can also
dominate, but this is the case for quite smooth asperities only and
for oscillations of very large azimuth indices. One can conclude
from Eqs.\eqref{R(h)_estim-alt(fin)} and \eqref{R(s1s2)_nrm-estim}
that for the amplitude scattering to be dominant it is necessary,
first, that the average tangent of the asperity angle of slope obey
the condition
\begin{subequations}\label{Height_dominate}
  \begin{equation}\label{R(h)>R(s2)}
   \overline{\tan\theta}\sim\frac{\sigma}{s_c}\ll
   \left(\frac{\sigma}{R}\right)^{1/2}\ .
  \end{equation}
Secondly, the inequality must be fulfilled
$\left(K_{\perp}R\right)^2\gg \Xi^2\left(R/s_c\right)^2$, which is
admissible for oscillations with quite large mode indices, i.\,e.,
  \begin{equation}\label{R(h)>R(s1)}
   n\gg\left(\frac{R}{s_c}\right)^2\ .
  \end{equation}
\end{subequations}
If only one of the inequalities \eqref{Height_dominate} fails to
hold (indeed, this is the case for the overwhelming majority of
parameters from the above-mentioned set), the gradient scattering
appears to be dominating.

%=======================================
\subsection{Whispering gallery modes of the rough-side DDR}
%=======================================

Based on the above estimations, equation \eqref{Psi_n-separate},
that governs the azimuth mode spectrum of the resonance system under
study can be substantially simplified if one considers the limiting
cases of weak and strong scattering. As is seen from
Eq.~\eqref{Vn-norm_estim-fin}, for small r.m.s. height of the
asperities (in the sense of inequality \eqref{SmallHeight}) the
intra-mode scattering due to potential $\widetilde{\mathcal{V}}_{n}$
is weak, thereby making it possible to disregard it in the main
approximation.

As far as the inter-mode scattering is concerned, it is not
straightforward to estimate it in the same way as the intra-mode
one. From Eqs.~\eqref{R(h)_estim-alt(fin)} and
\eqref{R(s1s2)_nrm-estim} it follows that inter-mode scattering
should be classified either as weak or strong one depending upon
whether the operator $\hat{\mathsf{R}}$ norm is small or large as
compared with unity. If the inter-mode scattering resulting from
both the amplitude and the gradient potentials is thought of as weak
one, which occurs when two inequalities hold simultaneously
\begin{subequations}\label{Weak_intermode_scatt}
  \begin{gather}
 \label{Weak_intermode(h)}
    k\sigma \ll 1\ ,\\[6pt]
 \label{Weak_intermode(s1)}
    \frac{\sigma}{s_c} \ll \frac{s_c}{R}\ ,
  \end{gather}
\end{subequations}
the potential $\hat{\mathcal T}_n$ in Eq.~\eqref{Psi_n-separate} can
be ignored with parametric accuracy. In this instance the resonator
spectrum can be obtained from the equation which differs from the
initial unperturbed one simply by renormalizing both the mode index
($n\to\widetilde{n}$) and the radial wave number
($K_{\perp}\to\widetilde{K}_{\perp}$). In the parameter region where
both of the inequalities \eqref{Weak_intermode_scatt} are satisfied
this renormalization is extremely small ($\Xi\ll\varphi_c$) and can
be safely disregarded.

Conditions \eqref{Weak_intermode_scatt}, that indicate the
inter-mode scattering weakness, can be easily violated. By virtue of
peculiar technology, when manufacturing micro- and nano-sized
quantum resonance systems it is normally difficult to obey, e.\,g.,
inequality \eqref{Weak_intermode(s1)} which corresponds to extreme
smoothness of surface inhomogeneities. Moreover, inequality
\eqref{Weak_intermode(h)} is fulfilled within only limited, i.\,e.,
long-wavelength, part of the resonator bandwidth.

If, at least, one of the conditions \eqref{Weak_intermode_scatt} is
violated, the inter-mode scattering fails to be classified as a weak
one in the sense that the norm of scattering operator
$\hat{\mathsf{R}}$ in the potential \eqref{T_oper-suppl} becomes
large as against the unity. Nevertheless, in this case one may as
well simplify the \emph{T}-potential by expanding it in a series in
the inverse scattering operator, $\hat{\mathsf{R}}^{-1}$. The
expression between the projection operators in
Eq.~\eqref{T_oper-suppl} is identically transformed as
\begin{align}
  \hat{\mathcal U}
  \left(\openone-\hat{\mathsf{R}}\right)^{-1}\hat{\mathsf{R}}= &
  -\hat{\mathcal U}+\hat{\mathcal
  U}\left(\openone-\hat{\mathsf{R}}\right)^{-1}=
  -\hat{\mathcal U}-\hat{\mathcal U}\hat{\mathsf{R}}^{-1}
  \left(\openone-\hat{\mathsf{R}}^{-1}\right)^{-1} \notag\\
\label{T_matr-large_R}
  =& -\hat{\mathcal U}-\hat{\mathcal U}\,\hat{\mathcal U}^{-1}
  \hat{\mathcal{G}}^{(V)-1}\left(\openone-\hat{\mathsf{R}}^{-1}\right)^{-1}
  \approx -\hat{\mathcal U}-\hat{\mathcal{G}}^{(V)-1}-
  \hat{\mathcal{G}}^{(V)-1}\hat{\mathcal
  U}^{-1}\hat{\mathcal{G}}^{(V)-1}\ .
\end{align}
Here we use the symbolic notation $\hat{\mathcal{G}}^{(V)-1}$ for
the full (i.\,e., including the whole set of azimuth modes) Green
operator with no intermode potentials. As a result of ``coating''
Eq.~\eqref{T_matr-large_R} with the projection operators $\bm{P}_n$
the first term in its right-hand side is eliminated, and the
operator acting on the wave function in Eq.~\eqref{Psi_n-separate}
assumes the following approximate form,
\begin{align}
  \frac{1}{r}\frac{\partial}{\partial
  r}r\frac{\partial}{\partial r}
  +\widetilde{K}_{\perp}^2(r|z_{\scriptscriptstyle\lessgtr}) & -\frac{\widetilde{n}^2}{r^2}
  -\widetilde{\mathcal{V}}_{n}(r|z_{\scriptscriptstyle\lessgtr})-\hat{\mathcal T}_n=
  \hat{G}_n^{(V)-1}-\hat{\mathcal T}_n
  \notag\\
\label{Diff_oper(R>>1)}
 & \approx 2\hat{G}_n^{(V)-1}+\bm{P}_n\hat{\mathcal{G}}^{(V)-1}\hat{\mathsf{R}}^{-1}\bm{P}_n
  \approx 2\left[\frac{1}{r}\frac{\partial}{\partial
  r}r\frac{\partial}{\partial r}
  +\widetilde{K}_{\perp}^2(r|z_{\scriptscriptstyle\lessgtr})-\frac{\widetilde{n}^2}{r^2}\right]\ .
\end{align}

By comparing the expression in the right-hand side of
Eq.~\eqref{Diff_oper(R>>1)} with the one in its left-hand side, with
potentials
$\widetilde{\mathcal{V}}_{n}(r|z_{\scriptscriptstyle\lessgtr})$ and
$\hat{\mathcal T}_n$ set to zero, one can notice that the limiting
cases of weak and strong (in the sense of the operator
$\hat{\mathsf{R}}$ norm) inter-mode scattering differ from one
another solely by doubling the wave operator in the latter case.
Clearly, such doubling implies that in strong surface scattering the
amplitude of the excited oscillations is twice as small as the one
in the weak scattering limit. However, it is evident that the change
in the common factor multiplying the wave operator cannot reveal
itself in dispersion relations.

We thus have demonstrated that the formal algebraic structure of the
wave operator remains basically unchanged under conditions of weak
and strong scattering. In both of these cases the wave operator
differs slightly from its unperturbed form. The main difference
between dispersion relations for resonators with perfect and rough
side boundaries is that in the latter case the initial transverse
wave parameter ${K}_{\perp}(r|z)$ and the mode index $n$ are
renormalized by the gradient factor of $(1+\Xi^2)^{-1/2}$. This
enables us to immediately write down the dispersion equations for a
DDR with random inhomogeneous side boundary based upon the results
given in Appendix~\ref{Unperturbed_spectrum}. Specifically, for a
rough-bounded DDR the oscillation spectrum is governed by the
equation
\begin{align}\label{Main_disp_eq-rough}
  \left[\frac{\varepsilon}{\widetilde{k}_{\perp}^{\varepsilon}}\cdot
  \frac{{J_{\widetilde{n}}}'(\widetilde{k}_{\perp}^{\varepsilon}R)}
  {J_{\widetilde{n}}(\widetilde{k}_{\perp}^{\varepsilon}R)}-
  \frac{1}{\widetilde{k}_{\perp}}\cdot\frac{{H_{\widetilde{n}}^{(1)}}'(\widetilde{k}_{\perp}R)}
  {H_{\widetilde{n}}^{(1)}(\widetilde{k}_{\perp}R)}\right]
  \Bigg[\frac{1}{\widetilde{k}_{\perp}^{\varepsilon}}\cdot
  \frac{{J_{\widetilde{n}}}'(\widetilde{k}_{\perp}^{\varepsilon}R)}
  {J_{\widetilde{n}}(\widetilde{k}_{\perp}^{\varepsilon}R)} -
  \frac{1}{\widetilde{k}_{\perp}}\cdot\frac{{H_{\widetilde{n}}^{(1)}}'(\widetilde{k}_{\perp}R)}
  {H_{\widetilde{n}}^{(1)}(\widetilde{k}_{\perp}R)}\Bigg]=
  \frac{\widetilde{n}^2}{R^2}(\varepsilon-1)^2
  \frac{k_z^2\widetilde{k}^2}{{(\widetilde{k}_{\perp}^{\varepsilon})}^4\widetilde{k}_{\perp}^4}\
  ,
\end{align}
which has to be supplemented with additional relationships between
$k$ and $k_z$, the latter resulting from joining the fields at the
end interfaces. In Eq.~\eqref{Main_disp_eq-rough}, the notations are
introduced
$\big(\widetilde{n},\widetilde{k},\widetilde{k}_{\perp}^{\varepsilon},\widetilde{k}_{\perp}\big)=
\big(n,k,k_{\perp}^{\varepsilon},k_{\perp}\big)\big/\sqrt{1+\Xi^2}$,
the wave numbers  $k_{\perp}^{\varepsilon}$ and $k_{\perp}$ are
defined in Eq.~\eqref{Wave_parameters}.

As the additional connection between wave parameters $k$ and $k_z$
in the case of a perfect cylindrical resonator
equation~\eqref{k_kz-sim} is obtained for $E_z$-symmetric
oscillations. For $E_z$-antisymmetric oscillations
Eq.~\eqref{k_kz-asim} holds. One can easily check that in going from
unperturbed wave equation \eqref{Psi_eq-unperturbed} to equation
\eqref{PsiEq-alt} describing the DDR with rough side wall the
equations derived through joining EM field components at the end
boundaries of the dielectric disk remain unchanged.

Finally, proceeding from the above calculations we are led to
conclude that with any roughness of the DDR side wall, which is
basically restricted by the smallness condition \eqref{SmallHeight},
to obtain the oscillation spectrum one can make use of the
relationship Eq.~\eqref{Main_disp_eq-rough} mainly coincident in
form with the dispersion equation for an infinitely long dielectric
cylinder. As a~supplementary condition to interconnect the
lengthwise and transverse wave vector components equation
\eqref{k_kz-sim} or \eqref{k_kz-asim} should be applied, depending
upon $E_z$-symmetry of the desired solution. The fundamental
difference of Eq.~\eqref{Main_disp_eq-rough} from its
perfect-cylinder counterpart is the renormalization of basic wave
parameters, which is governed by geometric properties of the
asperities. A special emphasis should be placed on the fact that it
is exactly the gradient scattering mechanism, not the amplitude one,
that most crucially affects the DDR spectrum. This particular fact
makes itself evident in that the renormalization of wave parameters
in Eq.~\eqref{Main_disp_eq-rough} is dictated not by the height-type
parameters, such as, e.\,g., Rayleigh parameter $k\sigma$ or the
ratio $\sigma/R$, but is mainly regulated by the mean-square slope
of the asperities against the unperturbed resonator boundary, which
is specified by parameter $\sigma/s_c$.

%=======================================
\section{Experimental results and discussion}
\label{Experiment}
%=======================================

The main goal of experimental studies in this work was to validate
our theory as regards microresonator spectra and the effect  of
random surface inhomogeneities on them. The point is that essential
assumption adopted in the theory is that it virtually does not
consider electromagnetic fields radiated from the resonator external
edges into the corner regions labeled by numbers 4 and 4$'$ in
Fig.~\ref{fig5} (see Appendix~\ref{Unperturbed_spectrum}). At the
same time, without making quite complex calculations one cannot make
\emph{a priori} statements about these fields being small enough to
neglect them in calculating the resonator spectrum. Yet another goal
of the experiment was to examine our theoretical findings concerning
the physical mechanism that adequately describes the influence of
random surface inhomogeneities on microresonator spectral
properties.

The studies of the microresonator spectrum were performed through
modeling these quite small systems in the millimeter wave band. For
this purpose we used a quasioptic dielectric disk resonator.
Physically, oscillation properties of this macroscopic system are
identical with properties of silicon microresonators used as
oscillation systems in real optical lasers. Our resonator was made
of teflon whose permittivity is not significantly far from unity
($\varepsilon=2.08$) and whose dielectric loss in the millimeter
range are fairly small ($\tan\delta\cong 2.3 \times 10^{-4}$). In
the model DDR, whispering gallery oscillations were excited with the
EM field concentrated at the periphery of the disk, in the narrow
region close to its side boundary. This enabled us to use the disk
core to fix it in the level position without introducing additional
dissipative losses to the experiment. The source of WG modes was
positioned close to the resonator side boundary. Its role was played
by the waveguide antenna powered by the microwave generator. The
antenna was fabricated as a waveguide tapered along the short wall,
whose butt end was positioned near the resonator side surface. The
receiving antenna was made identical to the source antenna and was
placed at the diametrically opposite point off the disk.

Thin dielectric bracket-bars were used as the inhomogeneities of the
resonator side boundary. They were attached to the DDR side surface.
The basic requirement imposed on the inhomogeneities was for them
not to cause noticeable additional dissipative losses. To this end,
the bracket-bars
\begin{figure}[h!!]
  \setcaptionmargin{.8in}%
  \centering \scalebox{.75}[.75]{\includegraphics{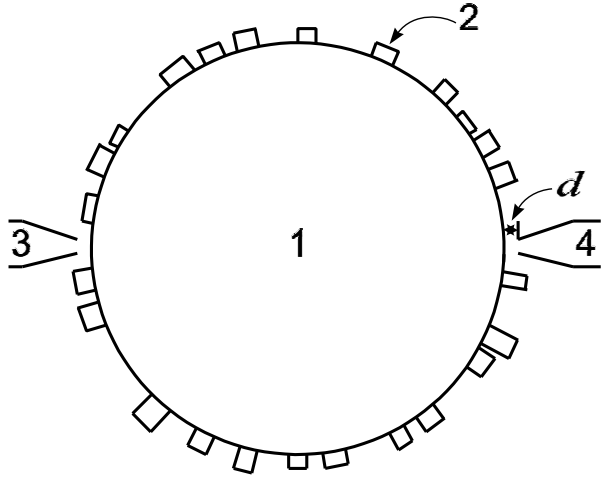}}
  \caption{Schematic view of the experimental DDR
  with surface inhomogeneities: 1 --- teflon disk; 2~--- superimposed teflon
  bracket-bars (inhomogeneities); 3, 4 --- the exciting and
  receiving waveguide antennas; $d$ --- the gap between
  the antenna and the DDR's side surface, which is used to tune
  the coupling of antenna with the resonator. The diameter of the DDR is 102~mm, the
  thickness is 7.6~mm. The inhomogeneities dimensions are:
  the~length, the width, and the thickness
  7.6~mm, 3~mm, and 2~mm, respectively.
  \hfill\label{fig2}}
\end{figure}
were made of the same teflon as the resonator body. The
inhomogeneities distribution on the resonator side surface was
random and varied for each different realization. In
Fig.~\ref{fig2}, the general view of the resonator is shown along
with the exciting and the receiving antennas as well as with the
attached teflon bracket-bars.

In order to excite TE or TM oscillations in the resonator, two
different configurations of antenna-versus-resonator were used. For
TE oscillations the antenna magnetic field was directed along the
resonator axis, $z$. To achieve this, the wide plate of the
waveguide was aligned parallel to this axis. In the case of TM
oscillations, along the axis $z$ the electric field was directed.
For this purpose the wide side of the antenna was oriented
transversely.

By changing the distance between the exciting waveguide butt end and
the resonator side boundary, as well as the angle between them, we
were able to adjust the coupling between the antenna and the
resonator to optimize it. As the optimal coupling we accepted the
one whereby the additional loss caused by the antenna was much less
than the eigen-loss in the resonator. At the same time, it was
necessary to keep the level of the coupling sufficient for spectral
lines to be traceable. The pattern of whispering gallery EM fields
in the DDR in known to be very sensitive to the frequency variation.
Therefore the resonator-to-waveguide coupling is different for each
of the modes. Based upon this, we established the optimal coupling
for each of the spectral lines separately while carrying out the
measurements in a wide frequency range.

Spectral measurements with the model resonator were made in the
on-pass regime using millimeter waveband standing-wave ratio meter.
Since the experiments were conducted over a wide range of
frequencies and for large number of realizations of surface
inhomogeneities, all spectral measurements were rendered automatic.
The signal from the ratio meter was sent to the computer and
processed by means of the especially designed program to find both
the frequencies and the quality factors of resonance lines. The
accuracy of the measurements was 0.01\% and 5\% for the resonance
frequencies and the quality factors, respectively.

\begin{figure}[h!!]
  \centering
  \includegraphics[width=.45\linewidth]{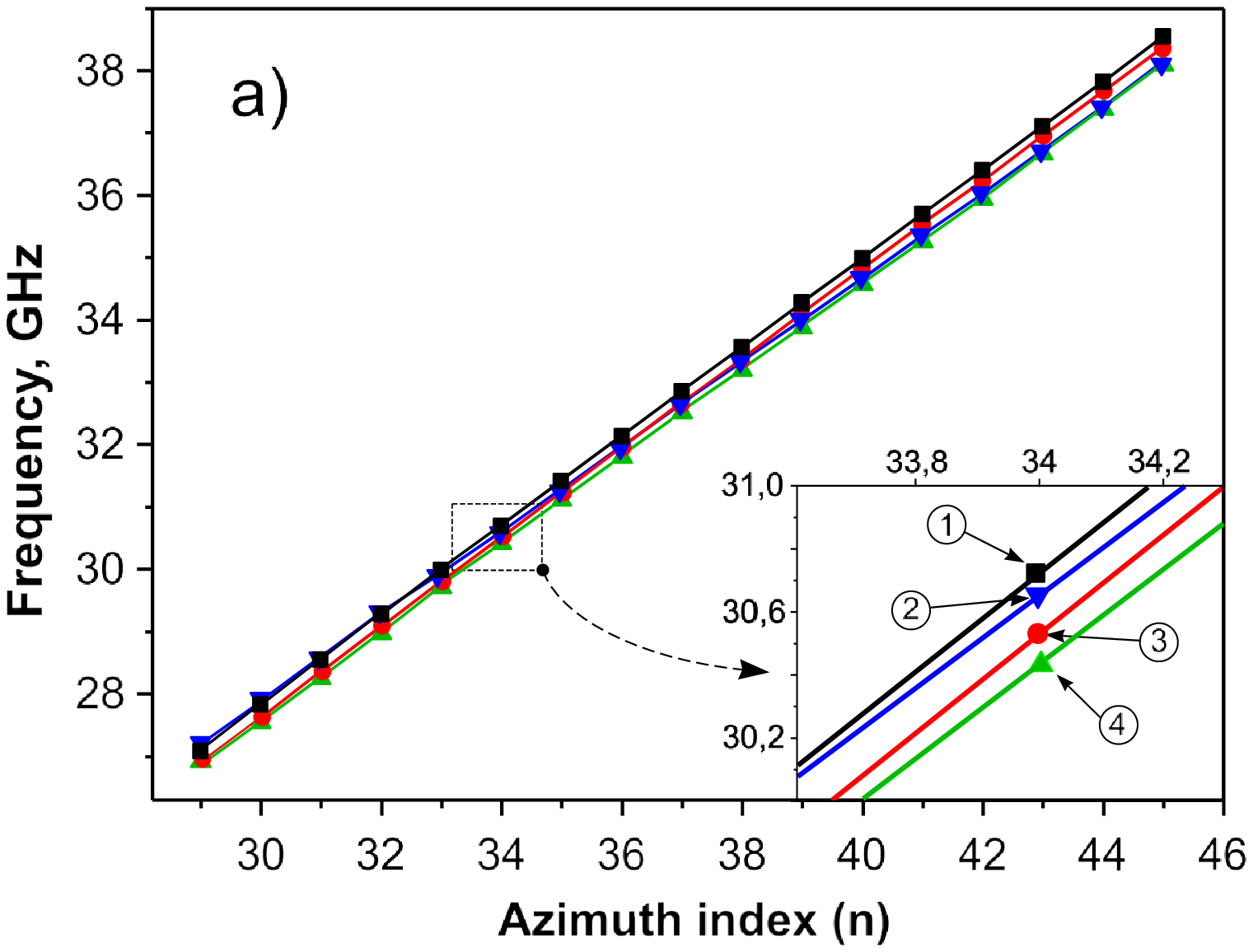}
  \hspace{1cm}
  \includegraphics[width=.45\linewidth]{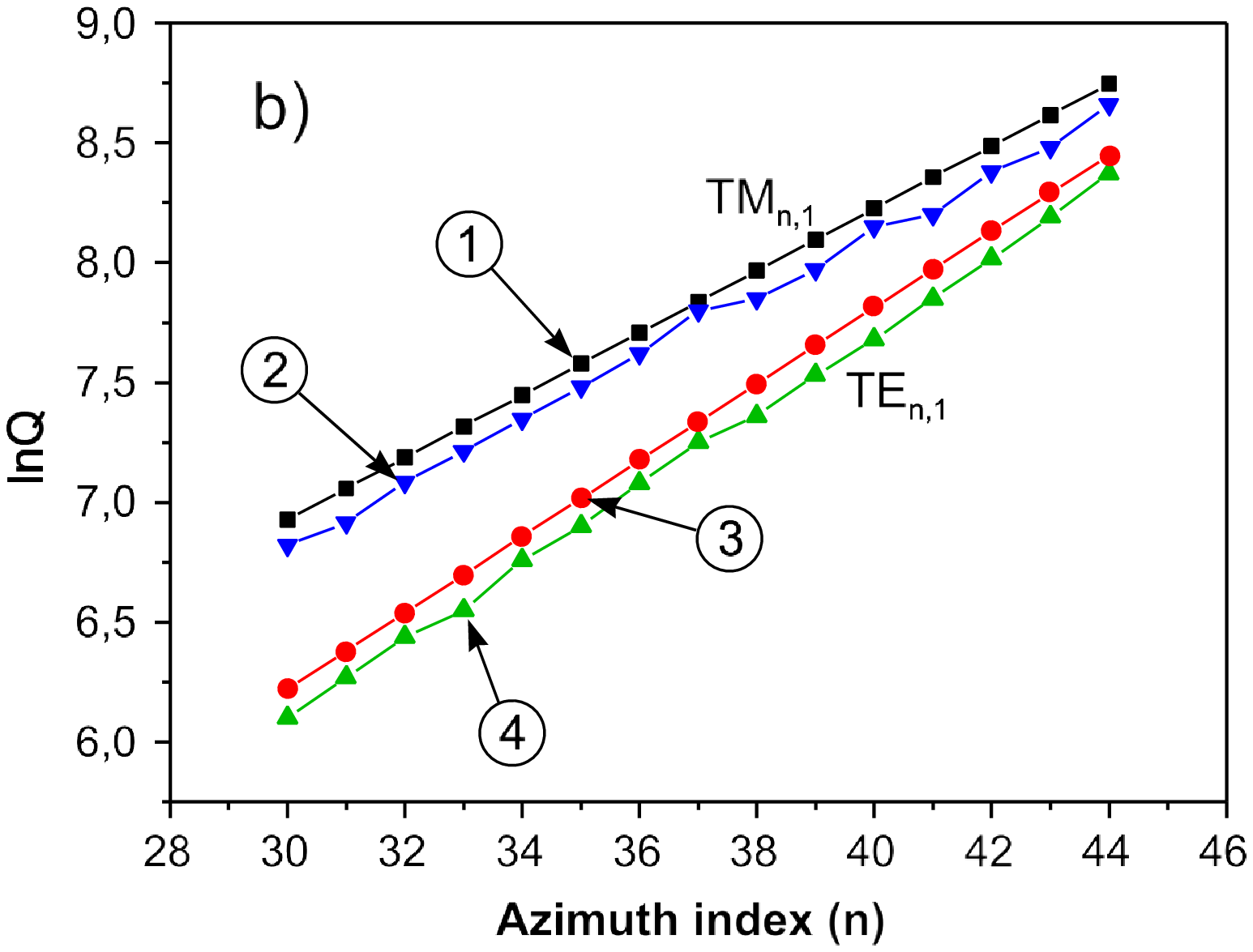}
  \caption{(Color online) The collation of theoretic and
  experimental data for the DDR with perfect cylindrical boundaries:
  the dependence a)~of spectral line frequency and b)~of the
  logarithm of quality factor on the azimuth index $n$.
  For $\mathrm{TM_{n,1}}$ mode: 1 ---calculation, 2 --- experiment;
  for $\mathrm{TE_{n,1}}$ mode: 3 --- calculation, 4 --- experiment.
  \hfill\label{fig3}}
\end{figure}

To identify the spectral lines, it was necessary to determine the
value of azimuth index for each of the lines in the absence of
inhomogeneities. To this end, the miniature rotating probe made of a
thin metal plate was used, which we inserted into the region at the
resonator disk where the electric field antinode was positioned. In
so doing, the source was tuned to the frequency of the particular
spectral line. When the probe rotated about the resonator axis, the
signal registered by the receiver varied in time at the rate the
probe passed across the regions with electric field loops. This
enabled us to determine the desired mode index through the
measurements of the signal modulation frequency. In Fig.~\ref{fig3},
the results of numerical calculations of the resonator spectrum are
shown along with spectral measurements data. As seen from
Fig.~\ref{fig3}a, the measured spectra of TE and TM modes correlate
well with the calculated spectra. The difference between spectral
line frequencies found from Eq.~\eqref{Main_disp_eq-rough},
including intrinsic
dissipation loss in the dielectric, and experimentally measured
frequencies is no more than 1\%. The quality factors obtained in the
experiment (as shown in Fig.~\ref{fig3}b) are smaller than those
calculated theoretically. We attribute this to the fact that the
measured $Q$-factor includes not only eigen-loss in the resonator
material but also is affected by the loss resulting from coupling
with the antenna. From the diagrams in Fig.~\ref{fig3} one can
observe the nearly equidistant character of the spectrum (the
frequency interval $\Delta\nu\cong 0.7$~GHz), which is typical for
open resonators with WG-type oscillations \cite{bib:Vainstein88}, as
well as the exponential dependence on the mode index~$n$ of the
calculated and the measured $Q$-factors.

Our calculations suggested, and that was experimentally confirmed,
that for TM oscillations the resonator quality factor is
significantly larger than that for TE oscillations, both of them
being taken with the same azimuth index. This suggests that in the
case of TM oscillations the DDR has the property to more efficiently
retain electromagnetic field in its volume than the resonator with
TE oscillations does. We have thus established that both the
theoretical model of the resonator we have chosen for our
calculations and the characteristic equations obtained thereupon
have found impressive experimental confirmations.

According to the findings of our theory, the main physical mechanism
for EM field scattering by random inhomogeneities of the resonator
side boundaries is the gradient mechanism. The scattering arisen due
to fluctuations in the asperity slope can be allowed for by way of
modification of the cylindrical functions indices along with the
wave numbers in the characteristic equation
Eq.~\eqref{Main_disp_eq-rough}. The modification reduces to
multiplying the parameters $n$, $k$, $k_{\perp}$ and
$k_{\perp}^{\varepsilon}$ by factor of
$\chi=\left(1+\Xi^2\right)^{-1/2}<1$. In order to identify the main
scattering mechanism experimentally and thus to check the developed
theory it was necessary to estimate the value of the parameter $\Xi$
starting from the parameters relevant to a particular experiment. We
were governed by the following considerations. Setting the
correlation length equal to an average distance between the centers
of the attached dielectric bracket-bars, i.\,e., $s_c=2\pi R/N$,
where $N$ is the total number of the bracket-bars, with Gaussian
distributed inhomogeneities we obtain
\begin{equation}\label{Xi-estim}
  \chi=\left[1+\left(\frac{\sigma N}{R}\right)^2\right]^{-1/2}\ .
\end{equation}
It can be easily seen that the parameter $\chi$ tends to decrease
with an increasing number of bracket-bars, which leads to a decrease
in the effective wave number $\widetilde{k}$ and the effective mode
index $\widetilde{n}$. Since the dependence of the quality factor on
the mode index is nearly exponential, $Q\approx
\mathrm{e}^{\widetilde{n}}$, the availability of surface
inhomogeneities should result in the decrease in the resonator
quality factor, whose origin is not an additional dissipative loss.

\begin{figure}[h!!]
  \setcaptionmargin{.4in}%
  \centering \scalebox{.6}[.58]{\includegraphics{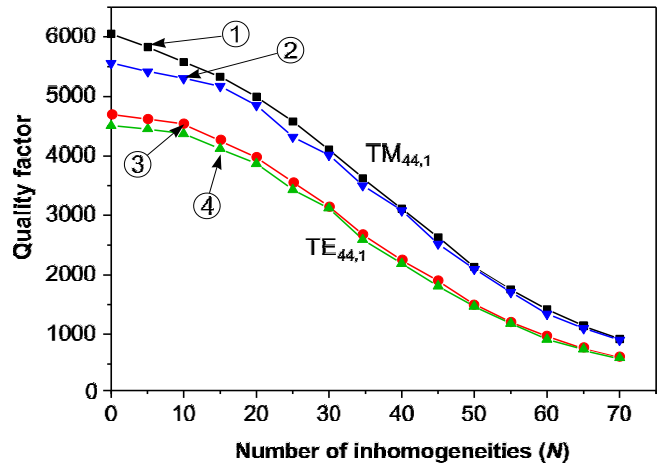}}
  \caption{(Color online) The $Q$-factor curves versus the
  number of inhomogeneities on the resonator side surface. $\mathrm{TM}_{44,1}$
  oscillations: 1 --- theory, 2 --- experiment; $\mathrm{TE}_{44,1}$
  oscillations; 3 --- theory, 4 --- experiment. Calculations were
  carried out at $\sigma=3.5$~mm and $R=51$~mm.
  \hfill\label{fig4}}
\end{figure}

According to our theory, with a small number of inhomogeneities the
effect of slope-controlled scattering must be fairly slight (the
parameter $\Xi$ decreases with lowering $N$). Therefore the value of
quality factor must be weakly dependent on $N$ as well. Such was
indeed the case in our experiments whose results are depicted in
Fig.~\ref{fig4}.
In the same figure, the curves $Q(N)$ are plotted for comparison,
which are calculated from characteristic equation
\eqref{Main_disp_eq-rough} including modification factor $\chi$ and
the dissipation loss in the resonator material. The loss was taken
into consideration phenomenologically, by adding the imaginary part
to permittivity $\varepsilon$, see Eq.~\eqref{Epsilon(r,z)}. With
the statistical nature of our measurements (the averaging is done
over a large number of realizations of the bracket-bar set), the
agreement between theoretical and experimental results appears to be
quite satisfactory. This suggests that, since both in our theory and
in the numerical calculations based upon it the slope scattering
mechanism alone is taken into account, the qualitative agreement
between theory and experiment unambiguously corroborates the
dominant role of this particular type of wave-surface scattering.

As is also seen from Fig.~\ref{fig4}, the influence of surface
inhomogeneities on the DDR spectrum differs significantly for TE-
and TM-type oscillations. The quality factors of TE oscillations are
to a far larger extent subjected to the resonator boundary roughness
as against the factors of TM oscillations. With a given number of
inhomogeneities, the $Q$-factor of TE oscillations is noticeably
smaller than that of TM oscillations. This fact, which is quite
important for practical applications of microresonators in laser
oscillation systems, was given certain attention in
Ref.~\cite{bib:Borselli2004}. Yet in the present work it has been
comprehensively substantiated.

%=======================================
\section{Conclusion}
\label{Conclusion}
%=======================================

To summarize, we have investigated spectral properties of a
dielectric disk resonator with randomly rough side boundary both
theoretically and experimentally. It is shown that the azimuth modes
of the resonator oscillations can be rigorously separated at
whatever level of surface inhomogeneities. This allowed us to obtain
the asymptotically exact dispersion equations which are valid over a
wide range of the roughness parameters. The only requirement imposed
on the inhomogeneities and effectively utilized in the calculations
was that their mean-square height be small as compared with the
unperturbed (i.\,e., non-rough) disk resonator radius.

In deriving dispersion relations, we have shown that electromagnetic
wave scattering resulting from the boundary roughness can be
described in terms of two fundamentally different physical
mechanisms, specifically, the amplitude (the height) scattering
mechanism and the gradient (the slope) mechanism. For the first of
them, the ratio of the mean height of the asperities to the
oscillation wavelength (the Rayleigh parameter) acts as the main
guiding parameter whereas for the gradient mechanism the mean slope
of the asperities relative to the unperturbed resonator boundary
plays the same role. Our estimations have revealed that it is
exactly the gradient scattering that is of primary importance for
the formation of rough resonator spectrum. The effect of this type
of scattering can be described through the approximate dispersion
equation that differs from the analogous equation for the ideally
circular DDR by the gradient renormalization of the basic wave
parameters, i.\,e,, the mode wave numbers and the azimuth index.

Our theory aimed at describing the influence of random surface
inhomogeneities on microresonator spectral properties analytically
is actually of model nature. The point is that it does not
incorporate the electromagnetic fields radiated from the resonator
into external angle sectors. To neglect these fields theoretically
is quite a challenge. Therefore, to verify the conclusions of our
approximate (to the certain extent) theory we have made experimental
measurements on the model millimeter wave band dielectric disk
resonator. Surface inhomogeneities were presented by teflon
bracket-bars of relatively small cross-section, which were attached
randomly to the resonator side boundary. For the perfect-wall
resonator, i.\,e. the resonator with no inhomogeneities attached,
our measurement data have demonstrated excellent agreement with the
developed theory concerning both the frequency spectrum and the
quality factors of spectral lines. Close qualitative agreement is
also attained with regard to the effect produced on the resonator
spectrum by random surface inhomogeneities placed at the side
boundary. First, we have confirmed the theoretical predictions about
the leading role of the gradient scattering mechanism in describing
the effect of random rough boundaries. Second, we ascertained that
the system of model dispersion equations we have obtained herein is
quite effective in describing the open microresonator spectra.

Besides, our experiments have revealed that the effect produced by
the surface inhomogeneities of the DDR on the TE and TM oscillation
spectra is fundamentally different. The TM oscillations quality
factor exceeds significantly the analogous factor for the TE
oscillations, being yet less affected by surface inhomogeneities. In
our work this particular fact, which is essential for the production
of microresonator-based lasers, has been theoretically and
experimentally validated.

\begin{acknowledgments}
This work was partially supported by the Science and Technology
Center of Ukraine (STCU), project No.~4114.
\end{acknowledgments}
%%%%%%%%%%%%%%%%%%%%%%%%%%%%%%%%%%%%%%%
\appendix
%%%%%%%%%%%%%%%%%%%%%%%%%%%%%%%%%%%%%%%

%=======================================
\section{Model spectrum of ideal cylindrical DDR}
\label{Unperturbed_spectrum}
%=======================================

Proceeding from the methodology considerations, we outline the
technique for deriving dispersion equations for cylindrical
finite-size resonator with perfectly smooth bounding surfaces.

Consider equation \eqref{GreenEqMain} without potentials
$\hat{V}^{(h)}$ and $\hat{V}^{(s)}$. Upon going to Fourier
representation over angle variable $\varphi$ using the complete set
of eigenfunctions $\ket{\varphi,n}=(2\pi)^{-1/2}\exp(-in\varphi)$,
where $n=0,\pm 1,\pm 2,\ldots\ $, the equation for $n$-th angular
component of the wave function becomes
\begin{equation}\label{Psi_eq-unperturbed}
  \left[\frac{1}{r}\frac{\partial}{\partial r}r
  \frac{\partial}{\partial r}+
  \frac{\partial^2}{\partial
  z^2}+K^2(r,z)-\frac{n^2}{r^2}\right]\Psi_n(r,z)=0\ .
\end{equation}
The function $\Psi_n(r,z)$ is sought to be finite as $r\rightarrow
0$, whereas at $r\rightarrow\infty$ and $|z|\to\infty$ the radiation
conditions are meant to be fulfilled.

It is difficult to find the explicit solution of
Eq.~\eqref{Psi_eq-unperturbed} in the entire domain of variables $r$
and $z$. Yet basically there is no need to make use of such a
solution. It will suffice to obtain dispersion relations which are
not strictly valid but satisfied with good accuracy. We will seek a
desired solution in the model form, as was done, e.\,g., in
Refs.~\cite{bib:IvanovKalin1988,bib:Peng96}, imposing the pair of
basic requirements. One of the requirements is to fulfil fundamental
boundary conditions at zero and infinite distances from the
resonator center, while the other is to provide correct joining of
the EM field components at the dielectric disk interfaces.

By representing the solution of Eq.~\eqref{Psi_eq-unperturbed} as a
sum of $E_z$-symmetric and $E_z$-antisymmetric summands (hereinafter
we will specify them by indices ($s$) and ($a$), respectively) we
will seek the Debye potentials in the form given below,
\begin{subequations}\label{Un(s)}
\begin{align}
\label{Un(s)-1}
 &\hspace{-1cm} U_{n1}^{(s)}(r,z)=A_n^{e(s)}J_n(k_{\perp}^{\varepsilon}r)
   \cos k_zz\ ,\\
\label{Un(s)-2}
 &\hspace{-1cm} U_{n2}^{(s)}(r,z)=B_n^{e(s)}H_n^{(1)}(k_{\perp}r)
   \cos k_zz\ ,\\
\label{Un(s)-3}
 &\hspace{-1cm} U_{n3}^{(s)}(r,z)=C_n^{e(s)}J_n(k_{\perp}^{\varepsilon}r)
   \exp(-\varkappa_z|z|)\ ,
\end{align}
\end{subequations}
\vspace{-\baselineskip}
\begin{subequations}\label{Vn(s)}
\begin{align}
\label{Vn(s)-1}
 & V_{n1}^{(s)}(r,z)=A_n^{m(s)}J_n(k_{\perp}^{\varepsilon}r)
   \sin k_zz\ ,\\
\label{Vn(s)-2}
 & V_{n2}^{(s)}(r,z)=B_n^{m(s)}H_n^{(1)}(k_{\perp}r)
   \sin k_zz\ ,\\
\label{Vn(s)-3}
 & V_{n3}^{(s)}(r,z)=C_n^{m(s)}J_n(k_{\perp}^{\varepsilon}r)
   \exp(-\varkappa_z|z|)\sgn z\ .
\end{align}
\end{subequations}
In Eqs.~\eqref{Un(s)} and \eqref{Vn(s)}, the subscripts 1, 2, 3
\begin{figure}[h!!]
  \setcaptionmargin{.8in}%
  \centering \scalebox{.7}[.7]{\includegraphics{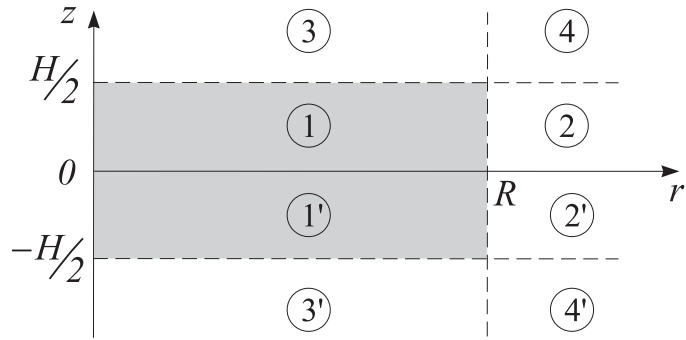}}
  \caption{The areas on the half-plane $(r,z)$ between
  which the joining of EM fields is performed. The shaded region is
  filled up with the dielectric.
  \hfill\label{fig5}}
\end{figure}
correspond to the regions (see Fig.~\ref{fig5}) which are labeled
with corresponding numbers. The wave parameters $k_{\perp}$,
$k_{\perp}^{\varepsilon}$ and $\varkappa_z$ are given by
\begin{subequations}\label{Wave_parameters}
\begin{eqnarray} \label{Wave_parameters-k_perp}
 && k_{\perp}^2 = k^2-k_z^2 \ ,\\
\label{Wave_parameters-ek_perp}
 &&{(k_{\perp}^{\varepsilon})}^2 = \varepsilon k^2-k_z^2\ ,\\
\label{Wave_parameters-kappa_z}
 && \varkappa_z^2 = k^2(\varepsilon-1)-k_z^2 \ .
\end{eqnarray}
\end{subequations}

By representing EM field components in terms of the potentials
\eqref{Un(s)} and \eqref{Vn(s)} and by joining the in- and the out-
field components at the resonator side boundary (i.\,e., between
regions 1 and 2) we obtain the well-known equation that couples
together wave parameters $k$ and $k_z$ \cite{bib:Vainstein88},
\begin{equation}\label{Main_disp_eq}
  \left[\frac{\varepsilon}{k_{\perp}^{\varepsilon}}
  \frac{{J_n}'(k_{\perp}^{\varepsilon}R)}{J_n(k_{\perp}^{\varepsilon}R)}-
  \frac{1}{k_{\perp}}\frac{{H_n^{(1)}}'(k_{\perp}R)}{H_n^{(1)}(k_{\perp}R)}\right]
  \Bigg[\frac{1}{k_{\perp}^{\varepsilon}}
  \frac{{J_n}'(k_{\perp}^{\varepsilon}R)}{J_n(k_{\perp}^{\varepsilon}R)}-
  \frac{1}{k_{\perp}}\frac{{H_n^{(1)}}'(k_{\perp}R)}{H_n^{(1)}(k_{\perp}R)}\Bigg]=
  \frac{n^2}{R^2}(\varepsilon-1)^2
  \frac{k_z^2k^2}{{(k_{\perp}^{\varepsilon})}^4k_{\perp}^4}\ .
\end{equation}
In order to obtain yet another coupling equation for the same wave
parameters we join the tangential components of the field at the end
boundary ${z=H/2}$. This leads to the set of equations
\begin{subequations}\label{End_matching}
\begin{align}
 & k_{\perp}^{\varepsilon}{J_n}'(k_{\perp}^{\varepsilon}r)
  \bigg[-k_z\sin\frac{k_zH}{2}A_n^{e(s)}  +
  \varkappa_z\exp\big(-\varkappa_zH/2\big)C_n^{e(s)}\bigg] \notag\\
 &\hspace{5cm} +\frac{n}{r}J_n(k_{\perp}^{\varepsilon}r)
  \,k\bigg[\sin\frac{k_zH}{2}A_n^{m(s)}-
  \exp\big(-\varkappa_zH/2\big)C_n^{m(s)}\bigg]=0\ ,\\
 & k_{\perp}^{\varepsilon}{J_n}'(k_{\perp}^{\varepsilon}r)
  \,k\bigg[\sin\frac{k_zH}{2}A_n^{m(s)} -
  \exp\big(-\varkappa_zH/2\big)C_n^{m(s)}\bigg]\notag\\
 &\hspace{5cm} +\frac{n}{r}J_n(k_{\perp}^{\varepsilon}r)
  \bigg[-k_z\sin\frac{k_zH}{2}A_n^{e(s)}  +
  \varkappa_z\exp\big(-\varkappa_zH/2\big)C_n^{e(s)}\bigg]=0\ , \\
 & k_{\perp}^{\varepsilon}{J_n}'(k_{\perp}^{\varepsilon}r)
  \bigg[k_z\cos\frac{k_zH}{2}A_n^{m(s)}  +
  \varkappa_z\exp\big(-\varkappa_zH/2\big)C_n^{m(s)}\bigg] \notag\\
 &\hspace{5cm} -\frac{n}{r}J_n(k_{\perp}^{\varepsilon}r)
  \,k\bigg[\varepsilon \cos\frac{k_zH}{2}A_n^{e(s)} -
  \exp\big(-\varkappa_zH/2\big)C_n^{e(s)}\bigg]=0\ ,\\
 & k_{\perp}^{\varepsilon}{J_n}'(k_{\perp}^{\varepsilon}r)
  \,k\bigg[\varepsilon \cos\frac{k_zH}{2}A_n^{e(s)} -
  \exp\big(-\varkappa_zH/2\big)C_n^{e(s)}\bigg] \notag\\
 &\hspace{5cm} -\frac{n}{r}J_n(k_{\perp}^{\varepsilon}r)
  \bigg[k_z\cos\frac{k_zH}{2}A_n^{m(s)}  +
  \varkappa_z\exp\big(-\varkappa_zH/2\big)C_n^{m(s)}\bigg]=0\ ,
\end{align}
\end{subequations}
where the number of the unknowns (to the latter we assign not only
constant factors included in Eqs.~\eqref{Un(s)} and \eqref{Vn(s)})
but also the functions ${J_n}'(k_{\perp}^{\varepsilon}r)$ and
$(n/r)J_n(k_{\perp}^{\varepsilon}r)$) exceeds the number of the
equations thus obtained. The situation can be improved if one adds
to the system \eqref{End_matching} a pair of equalities resulting
from joining the EM field normal components at the same end
boundary, specifically,
\begin{subequations}\label{Norm_comp_EH}
\begin{align}
   A_n^{e(s)}\cos\frac{k_zH}{2} & =
   \frac{1}{\varepsilon}C_n^{e(s)}\exp\big(-\varkappa_zH/2\big)\ ,\\
   A_n^{m(s)}\sin\frac{k_zH}{2} & =
   C_n^{m(s)}\exp\big(-\varkappa_zH/2\big)\ .
\end{align}
\end{subequations}
Using Eqs.~\eqref{End_matching} and \eqref{Norm_comp_EH} we arrive
at a set of four coupled equations which can be naturally combined
in pairs, viz.
\begin{subequations}\label{EesEms-eqs}
\begin{align}
 \label{EesEms-eqs1}
  & \left\{
   \begin{aligned}
    k_{\perp}^{\varepsilon}{J_n}'(k_{\perp}^{\varepsilon}r)
    \left(\varkappa_z-\frac{k_z}{\varepsilon}\tan\frac{k_zH}{2}\right)
    C_n^{e(s)}=0\ ,\\
    \frac{n}{r}J_n(k_{\perp}^{\varepsilon}r)
    \left(\varkappa_z-\frac{k_z}{\varepsilon}\tan\frac{k_zH}{2}\right)
    C_n^{e(s)}=0\ ,
   \end{aligned} \right.
 \\
 \label{EesEms-eqs2}
   & \left\{
   \begin{aligned}
    k_{\perp}^{\varepsilon}{J_n}'(k_{\perp}^{\varepsilon}r)
    \left(\varkappa_z+k_z\cot\frac{k_zH}{2}\right)
    C_n^{m(s)}=0\ ,\\
    \frac{n}{r}J_n(k_{\perp}^{\varepsilon}r)
    \left(\varkappa_z+k_z\cot\frac{k_zH}{2}\right)
    C_n^{m(s)}=0\ .
   \end{aligned} \right.
\end{align}
\end{subequations}

Since the Bessel functions do not vanish simultaneously with their
derivatives, one can satisfy Eqs.~\eqref{EesEms-eqs} in two ways.
The first one is to equate the expression in parenthesis of
Eq.~\eqref{EesEms-eqs1} to zero, putting the constant
${C_n^{m(s)}=0}$ in Eqs.~\eqref{EesEms-eqs2}. Alternatively, the
parenthesis in Eq.~\eqref{EesEms-eqs2} can be set equal to zero
concurrently with the coefficient $C_n^{e(s)}$ in
Eqs.~\eqref{EesEms-eqs1}. As can be readily seen from
Eqs.~\eqref{Un(s)} and \eqref{Vn(s)}, the former solution
corresponds to the TM-type oscillations whereas the latter is
consistent with TE polarization. Both of these cases can be combined
into one equation of the following form
\begin{equation}\label{k_kz-sim}
  \bigg(\underbrace{\varkappa_z-\frac{k_z}%
  {\varepsilon}\tan\frac{k_zH}{2}}_{\mathrm{TM}}\bigg)
  \bigg(\underbrace{\varkappa_z+k_z\cot\frac{k_zH}{2}}_{\mathrm{TE}}\bigg)=0
  \ .
\end{equation}
The lower curly brackets in Eq.~\eqref{k_kz-sim} indicate the
polarization which correlates with vanishing the expression in the
corresponding parenthesis. By carrying out the calculations similar
to those given above, for $E_z$-antisymmetric solution of
Eq.~\eqref{Psi_eq-unperturbed} we arrive, instead of
Eq.~\eqref{k_kz-sim}, at the equality
\begin{equation}\label{k_kz-asim}
  \bigg(\underbrace{\varkappa_z-k_z\tan\frac{k_zH}{2}}_{\mathrm{TE}}\bigg)
  \bigg(\underbrace{\varkappa_z+\frac{k_z}%
  {\varepsilon}\cot\frac{k_zH}{2}}_{\mathrm{TM}}\bigg)=0\ .
\end{equation}

Note that the relationships \eqref{Main_disp_eq}, \eqref{k_kz-sim}
and \eqref{k_kz-asim} were previously obtained in
Ref.~\cite{bib:IvanovKalin1988}, although the latter two of them
were of somewhat different form. In that paper, however,
polarization of the excited EM field was not discussed at all. At
first glance, from our derivation of Eqs.~\eqref{k_kz-sim} and
\eqref{k_kz-asim} it may seem that TM- and TE-type oscillations in
the DDR of finite thickness can be perfectly separated, at least,
with no regard for the boundary inhomogeneity. As a matter of fact,
this is not the case because one must take into account the
inaccuracy, i.\,e., the model character of wave solutions
\eqref{Un(s)} and \eqref{Vn(s)} used when deriving dispersion
relations. These solutions, as well as the analogous ones for the
antisymmetric case, are well adapted for joining the in- and the
out- field components at the boundaries between regions 1--2 and
1--3 in Fig.~\ref{fig5}. The fields in region 4 are not taken into
consideration, which must inevitably result in some ``overflow'' of
the spectrum obtained in such a way.

It is a difficult task to correctly evaluate the possibility to
neglect the fields in the 4-th region in Fig.~\ref{fig5} without
resorting to rigorous calculations. Nevertheless, by now there does
not exist a rigorous theory for a finite thickness DDR, even in the
seemingly simple case where the boundaries are perfectly smooth. For
this reason, in order to test the results we obtained by the model
calculations we choose to compare the spectrum resulting from
Eqs.~\eqref{Main_disp_eq-rough}, \eqref{k_kz-sim}, \eqref{k_kz-asim}
with the one measured in the experiment.

%=======================================
\section{Mode separation in two-dimensional wave equation with
arbitrary scattering potential}%
\label{Mode_separation}
%=======================================

Consider the equation for Green function of
Eq.~\eqref{Psi-Azim_repr} without rendering both the physical nature
of mode potentials and their absolute value concrete,
\begin{equation}\label{G-Azim_repr}
  \left[\frac{1}{r}\frac{\partial}{\partial r}r\frac{\partial}{\partial r}
  +\widetilde{K}_{\perp}^2(r|z_{\scriptscriptstyle\lessgtr})-\frac{\widetilde{n}^2}{r^2}
  -\widetilde{\mathcal{V}}_{n}(r|z_{\scriptscriptstyle\lessgtr})\right]
  G_{nn'}(r,r'|z)-\sum_{m\neq n}\widetilde{\mathcal{U}}_{nm}(r|z_{\scriptscriptstyle\lessgtr})
  G_{mn'}(r,r'|z_{\scriptscriptstyle\lessgtr})=
  \frac{1}{r}\delta(r-r')\delta_{nn'}\ .
\end{equation}
Along with the exact Green function, which has a matrix structure in
variables $n$ and $r$, we introduce for each of the azimuth modes
the trial mode propagator, which is assumed to obey the closed
equation
\begin{equation}\label{Trial_G-alt}
  \left[\frac{1}{r}\frac{\partial}{\partial r}r\frac{\partial}{\partial r}
  +\widetilde{K}_{\perp}^2(r|z_{\scriptscriptstyle\lessgtr})-\frac{\widetilde{n}^2}{r^2}
  -\widetilde{\mathcal{V}}_{n}(r|z_{\scriptscriptstyle\lessgtr})\right]
  G^{(V)}_{n}(r,r'|z_{\scriptscriptstyle\lessgtr})=\frac{1}{r}\delta(r-r')
\end{equation}
resulting from Eq.~\eqref{G-Azim_repr} providing that the inter-mode
scattering is disregarded. We will use the notation
$\hat{G}^{(V)-1}_{n}$ for the operator in square brackets of
Eq.~\eqref{Trial_G-alt}. The initial equation \eqref{G-Azim_repr}
can then be recast as
\begin{equation}\label{1D-trial-suppl}
  \hat{G}^{(V)-1}_{n}G_{nn'}(r,r'|z_{\scriptscriptstyle\lessgtr})=
  \frac{1}{r}\delta(r-r')\delta_{nn'}+
  \sum_{m\neq n}\widetilde{\mathcal{U}}_{nm}(r|z_{\scriptscriptstyle\lessgtr})
  G_{mn'}(r,r'|z_{\scriptscriptstyle\lessgtr})\ ,
\end{equation}
or, in the equivalent integral form, as
\begin{equation}\label{1D-trial-suppl-int}
  G_{nn'}(r,r'|z_{\scriptscriptstyle\lessgtr})=
  G_{n}^{(V)}(r,r'|z_{\scriptscriptstyle\lessgtr})\delta_{nn'}+
  \sum_{m\neq n}\int\limits_0^{\infty}r_1dr_1
  G_{n}^{(V)}(r,r_1|z_{\scriptscriptstyle\lessgtr})
  \widetilde{\mathcal{U}}_{nm}(r_1|z_{\scriptscriptstyle\lessgtr})
  G_{mn'}(r_1,r'|z_{\scriptscriptstyle\lessgtr})\ .
\end{equation}
By setting in Eq.~\eqref{1D-trial-suppl-int} index $n\neq n'$ and
then re-labelling all mode indices we can write this equation as the
equation for solely inter-mode components of the Green matrix
$\|G_{mn}\|$, the intra-mode ones being thought of as known
functions,
\begin{equation}\label{G_mn-integr_eq}
  G_{mn}(r,r'|z_{\scriptscriptstyle\lessgtr})-\sum_{\genfrac{}{}{0pt}{}{k\neq m}{k\neq n}}
  \int\limits_0^{\infty}r_1dr_1G_{m}^{(V)}(r,r_1|z_{\scriptscriptstyle\lessgtr})
  \widetilde{\mathcal{U}}_{mk}(r_1|z_{\scriptscriptstyle\lessgtr})
  G_{kn}(r_1,r'|z_{\scriptscriptstyle\lessgtr})=
  \int\limits_0^{\infty}r_1dr_1G_{m}^{(V)}(r,r_1|z_{\scriptscriptstyle\lessgtr})
  \widetilde{\mathcal{U}}_{mn}(r_1|z)G_{nn}(r_1,r'|z_{\scriptscriptstyle\lessgtr})\ .
\end{equation}

At this stage we introduce three operators,
$\hat{\mathcal{G}}^{(V)}$, $\hat{\mathcal U}$ and
$\hat{\mathsf{R}}$, which are assumed to act in the reduced
coordinate-mode space $\mathsf{\overline M}_n$ consisting of the
half-axis $r\geqslant 0$ and the entire set of mode indices except
for particular index $n$. The operators are specified by their
matrix elements
\begin{subequations}\label{G(V)_U-matr_elements-suppl}
\begin{align}
 \label{GV-matr_element-suppl}
  & \bra{r,n}\hat{\mathcal{G}}^{(V)}\ket{r',m} =
  G^{(V)}_{n}(r,r'|z_{\scriptscriptstyle\lessgtr})\delta_{nm}\ ,\\
 \label{U-matr_element-suppl}
  & \bra{r,n}\hat{\mathcal U}\ket{r',m}=\widetilde{\mathcal{U}}_{nm}(r|z_{\scriptscriptstyle\lessgtr})
  \frac{1}{r}\delta(r-r')\ ,\\
 \label{R-matr_element-suppl}
  & \bra{r,n}\hat{\mathsf{R}}\ket{r',m}=
  G^{(V)}_{n}(r,r'|z)\widetilde{\mathcal{U}}_{nm}(r'|z_{\scriptscriptstyle\lessgtr})\ .
\end{align}
\end{subequations}
Equation \eqref{G_mn-integr_eq} can now be recast as the matrix
equality
\begin{equation}\label{G_mn-integr_eq2}
  \bra{r,m}\big(1-\hat{\mathsf{R}}\big)\hat{\mathcal{G}}\ket{r',n}=
  \bra{r,m}\hat{\mathsf{R}}\bm{P}_n\hat{\mathcal{G}}\ket{r',n}\ ,
\end{equation}
or, equivalently, in the general operator form as
$\bm{P}_m\big(1-\hat{\mathsf{R}}\big)\hat{\mathcal{G}}\bm{P}_n=
\bm{P}_m\hat{\mathsf{R}}\bm{P}_n\hat{\mathcal{G}}\bm{P}_n$. Here,
$\bm{P}_n$ is the projection operator whose action reduces to
assigning the value $n$ to the nearest mode index of an arbitrary
operator standing adjacent to it, either to the left or right.
Multiplying both sides of the operator equality thus obtained by the
operator $\big(1-\hat{\mathsf{R}}\big)^{-1}$ whose regularity was
substantiated in Ref.~\cite{bib:Tar00} we arrive at the integral
relation between the non-diagonal and diagonal mode matrix elements
of the Green function, viz.
\begin{equation}\label{Gmn->Gnn-suppl}
  G_{mn}(r,r'|z_{\scriptscriptstyle\lessgtr})=
  \bra{r,m}\big(1-\hat{\mathsf{R}}\big)^{-1}
  \hat{\mathsf{R}}\bm{P}_n\hat{\mathcal{G}}\ket{r',n}=
  \big(\hat{K}_{mn}\hat{G}_{nn}\big)(r,r'|z_{\scriptscriptstyle\lessgtr})\ .
\end{equation}
Setting in Eq.~\eqref{G-Azim_repr} index $n'=n$ and substituting the
inter-mode propagators in the form \eqref{Gmn->Gnn-suppl} we
eventually obtain the closed equation for intra-mode Green function
$G_{nn}(r,r'|z_{\scriptscriptstyle\lessgtr})$,
\begin{equation}\label{Gnn_eq-suppl}
  \left[\frac{1}{r}\frac{\partial}{\partial r}r\frac{\partial}{\partial r}
  +\widetilde{K}_{\perp}^2(r|z_{\scriptscriptstyle\lessgtr})-\frac{\widetilde{n}^2}{r^2}
  -\widetilde{\mathcal{V}}_{n}(r|z_{\scriptscriptstyle\lessgtr})-
  \hat{\mathcal T}_n\right]
  G_{nn}(r,r'|z_{\scriptscriptstyle\lessgtr})=\frac{1}{r}\delta(r-r')\ .
\end{equation}
Here, $\hat{\mathcal T}_n$ is the operator that accounts for the
inter-mode scattering (it is just the $T$-matrix well-known in
quantum scattering theory \cite{bib:Newton68,bib:Taylor72}), which
has the form
\begin{equation}\label{T_oper-suppl}
  \hat{\mathcal T}_n=\bm{P}_n\hat{\mathcal U}
  \big(\openone-\hat{\mathsf{R}}\big)^{-1}\hat{\mathsf{R}}\bm{P}_n\
  .
\end{equation}
Trial Green function
$G^{(V)}_{n}(r,r'|z_{\scriptscriptstyle\lessgtr})$ entering operator
potential \eqref{T_oper-suppl} through matrix elements
\eqref{R-matr_element-suppl} obeys equation \eqref{Trial_G-alt} and
the same boundary conditions just like the desired mode propagator
$G_{nn}(r,r'|z_{\scriptscriptstyle\lessgtr})$.

Equation \eqref{Gnn_eq-suppl} along with Eq.~\eqref{Gmn->Gnn-suppl}
determines the entire Green function of wave equation
\eqref{Psi-Azim_repr} and serves as a basis for deriving the system
of uncoupled equations \eqref{Psi_n-separate} for azimuth components
of the sought-for wave function.

%%%%%%%%%%%%%%%%%%%%%%%%%%%%%%%%%%%%%%%

%%%%%%%%%%%%%%%%%%%%%%%%%%%%%%%%%%%%%%%

\begin{thebibliography}{cc}

\bibitem{bib:Polman04}
A. Polman, B. Min, J.~Kalkman, T.\,J.~Kippenberg, and K.\,J.~Vahala,
Appl. Phys. Lett. \textbf{84}, 1037 (2004).

\bibitem{bib:Vahala03}
K.\,J.~Vahala, Nature \textbf{424}, 839 (2003).

\bibitem{bib:GanErTar07}
E.\,M. Ganapolskii, Z.\,E. Eremenko, and Yu.\,V. Tarasov, \pre
\textbf{75}, 026212 (2007).

\bibitem{bib:Little97}
B.\,E. Little, J.-P. Laine, and S.\,T. Chu, Opt. Lett. \textbf{22},
4 (1997).

\bibitem{bib:Gorod00}
M.\,L. Gorodetsky and A.\,D. Pryamikov, J. Opt. Soc. Am. B
\textbf{17}, 1051 (2000).

\bibitem{bib:Oraevsk2002}
A.\,N. Oraevsky, Quantum Electron.\textbf{32}(5), 377 (2002).

\bibitem{bib:Borselli2004}
M.\,Borselli, K.\,Srinivasan, P.\,E. Barclay, and O.\,Painter, Appl.
Phys. Lett. \textbf{85}, 3693 (2004).

\bibitem{bib:Kuznetsov83}
M. Kuznetsov and H.\,A. Haus, IEEE J. Quan. Elec. \textbf{19}, 1505
(1983).

\bibitem{bib:BassFuks79}
F.\,G. Bass and I.\,M. Fuks, {\it Wave Scattering from Statistically
Rough Surfaces} (New York: Pergamon, 1979)

\bibitem{bib:Ogilvy91}
J.\,A. Ogilvy, \emph{Theory of Wave Scattering from Random Rough
Surfaces} (IOP Pub., Bristol, UK, 1991).

\bibitem{bib:Voronovich94}
A.G. Voronovich, \emph{Wave Scattering from Rough Surfaces}
(Springer Verlag, 1994).

\bibitem{bib:Rayleigh1907}
Lord Rayleigh, Proc. R. Soc. London Ser. A \textbf{79}, 399 (1907).

\bibitem{bib:Rayleigh45}
Lord Rayleigh, \emph{The Theory of Sound}, (Dover, New-York, 1945).

\bibitem{bib:MakTar98}
N.\,M. Makarov and Yu.\,V. Tarasov, J. Phys.: Condens. Matter
\textbf{10}, 1523 (1998).

\bibitem{bib:MakTar01}
N.\,M. Makarov and Yu.\,V. Tarasov, \prb {\bf 64}, 235306 (2001).

\bibitem{bib:Vainstein88}
L. A. Vainshtein, \emph{Electromagnetic Waves} [in Russian] (Radio i
Svyaz, Moscow, 1988).

\bibitem{bib:IvanovKalin1988}
E.\,Ivanov and V.\,I. Kalinichev, Radiotechnika [in Russian],
No.~10, 86 (1988).

\bibitem{bib:Peng96}
H. Peng, IEEE Trans. Microwave Theory Tech. \textbf{44}, 848 (1996).

\bibitem{bib:Anino97}
G. Annillo, M. Cassettari, I. Longo, and M. Martinelli, Chem. Phys.
Lett. \textbf{281}, 306 (1997).

\bibitem{bib:Tar00}
Yu.\,V. Tarasov, \WRM {\bf 10}, 395 (2000).

\bibitem{bib:Tar03}
Yu.\,V. Tarasov, Low. Temp. Phys. \textbf{29}, 45 (2003).

\bibitem{bib:Tar05}
Yu.\,V. Tarasov, \prb \textbf{71}, 125112 (2005).

% \bibitem{bib:LifGredPast88}
% I.\,M. Lifshits, S.\,A.Gredeskul, L.\,A.Pastur. \emph{Introduction
% to the Theory of Disordered Systems} (Wiley, New York, 1988).

\bibitem{bib:KolmFom68}
A.\,N. Kolmogorov, S.\,V. Fomin . \textit{Elements of the Theory of
Functions and Functional Analysis} (Dover, New York, 1961).

\bibitem{bib:Kato66}
T. Kato. \textit{Perturbation Theory for Linear Operators}
(Springer, Berlin, 1966).

\bibitem{bib:Newton68}
R.~Newton. {\it Scattering Theory of Waves and Particles}
(McGraw-Hill, New York, 1968).

\bibitem{bib:Taylor72}
J.\,R.~Taylor. {\it Scattering Theory. The Quantum Theory on
Nonrelativistic Collisions} (Wiley, New York, 1972).

\end{thebibliography}
\end{document}